# Main Manuscript for

## The crystalline properties of silica-witherite biomorphs vary within and between morphologies

*Moritz P.K. Frewein[1‡], Britta Maier[2‡], Moritz L. Stammer[1], Isabella Silva Barreto[1], Anastasiia Sadetskaia[3], Asma Medjahed[4], Remi Tucoulou[4], Julie Villanova[4], Manfred Burghammer[4], Henrik Birkedal[3], and Tilman A. Grünewald [1*]*

[1] Aix Marseille Univ, CNRS, Centrale Med, Institut Fresnel, Marseille, France
[2] University of Konstanz, Konstanz, Germany
[3] Department of Chemistry and INANO, Aarhus University, Aarhus, Denmark
[4] European Synchrotron Radiation Facility, Grenoble, France
[‡] These authors contributed equally.
[*] Corresponding author: Tilman A. Grünewald
**Email:** tilman.grunewald@fresnel.fr

**Author Contributions:**

M.P.K.F, B.M and T.A.G designed the research. M.P.K.F., B.M., M.L.S., I.S.B., A.M., R.T., J.V., M.B. and T.A.G carried out all experiments. M.P.K.F., A.S., H.B. and T.A.G analyzed the data. M.P.K.F., B.M, H.B. and T.A.G wrote the article.

**Competing Interest Statement:** The authors declare no competing interests.

**Classification:** Physical sciences - Chemistry

**This PDF file includes:**

Main Text
Figures 1 to 8
Table 1

## Abstract

Silica-witherite biomorphs are a class of emergent materials, i.e. composite microstructures made of nanometric barium carbonate surrounded by amorphous silica. They form via co-precipitation of barium carbonate and siliceous species, and self-organize into a multitude of shapes with a distinct long-range order of the carbonate nanocrystals. However, the internal structural organization within and across different morphologies remains insufficiently resolved.

Here, we use X-ray texture and diffraction tomography to create three-dimensional, spatially resolved maps of crystallographic orientation and structural parameters in silica-witherite biomorphs. At the sub-µm voxel level, all morphologies exhibit a crystallographic order consistent with a fiber texture around the c-axis. At larger length scales, however, the orientation field as well as crystallite size, crystallite shape anisotropy and the unit cell volume show systematic spatial variations. Leaf-like and helical morphologies contain defined directions along which structural parameters change systematically. Furthermore, we find recurring structural regimes with strong similarities between these morphologies. Conversely, coral-like morphologies are overall less textured and outside of the nucleation region we do not find clear structural regimes in the crystalline properties. These results provide a three-dimensional description of the internal organization of crystallites in silica–witherite biomorphs and establish a basis for systematically relating crystallographic organization to morphology.

**Significance Statement** Silica-witherite biomorphs are self-organizing microstructures of interest both as a functional material for the bottom-up fabrication of devices, and as a platform to study geological formation under extreme conditions. However, little is known about the crystalline properties of the carbonate phase, its spatial arrangement and crystalline texture. Using 3D crystallographic reconstructions of multiple silica-witherite biomorph morphologies, we reveal characteristic patterns of crystalline properties and orientation, as well as pronounced variations within single structures and across different morphologies. These nanoscale signatures reveal recurring structural regimes across morphologies. Our findings provide a structural basis for understanding biomorph morphogenesis and for guiding future functional-material design and *in situ* studies.

## Main Text

### Introduction

Silica-witherite biomorphs are self-organized inorganic microstructures formed by coupled carbonate-silicate precipitation under alkaline conditions (1–14). Due to their complex and hierarchical architectures, they serve as model systems for inorganic life-like morphogenesis (7, 9, 10) and as platforms for functional materials(3, 4, 11–14). Fig 1a-d show four of the most recurrent morphologies reported in literature.

These structures consist of anisotropic witherite ($BaCO_3$) nanocrystals embedded in an amorphous silica matrix (Fig 1e). Previous studies show that the nanocrystals typically exhibit a pseudo-hexagonal columnar morphology, with dimensions of ~200–400 nm in length and ~20–100 nm in width (15–19). Studies using birefringence, polarization microscopy, and computed tomography have further demonstrated that these nanocrystals show collective alignment and form orientation domains across larger regions of the biomorph. Variations in crystallite density and orientation contribute to mesoscale texture and optical anisotropy, highlighting the hierarchical coupling between nanoscale order and macroscopic structure (11, 17–19).

Biomorph morphology depends strongly on reaction conditions, with the balance between carbonate precipitation and the evolving silicate-rich growth environment cited as key factors (16, 22–25). Increasing sodium metasilicate concentration has been shown to not only alter macroscopic morphology, but also to reduce crystallite size, increase unit cell volume, and promote a transition from single-crystal to polycrystalline growth (15). A broader overview of the morphology space is presented in the morphograms in SI Fig. 1 along with a more detailed literature review on biomorph structure and chemistry. A study from Knoll and Steinbock further indicates that witherite nanocrystals evolve from dot-shaped precursors into elongated rods, with crystallographic alignment established rapidly after nucleation (18). We note, however, that this study relied mostly on surface-sensitive 2D methods. These previous observations demonstrate that crystallite size, anisotropy, and lattice parameters are coupled to local chemical conditions, suggesting that the internal crystalline organization of biomorphs encodes information about their growth environment. Such structure-property correlations are central for understanding how local chemistry, reaction-diffusion processes, and growth kinetics govern morphology, and provide a potential lever for structural optimization in functional materials.

Despite these advances, existing characterization approaches impose important limitations. Powder X-ray diffraction averages crystallographic information over large ensembles, electron microscopy probes only localized regions, and absorption-contrast X-ray tomography lacks sensitivity to crystallographic orientation and texture. While tomography has revealed internal density gradients and heterogeneous architectures, it cannot resolve the orientation or texture of the $BaCO_3$ nanocrystals. Consequently, the three-dimensional spatial distribution, orientation, and angular dispersion of nanocrystals within intact biomorphs remain largely unresolved.

This lack of spatially resolved crystallographic information limits our ability to relate nanoscale structure to growth dynamics and morphology. In particular, it remains unclear how crystalline properties evolve during growth and whether different morphologies share common structural regimes, restricting predictive control over biomorph formation.

Addressing these challenges requires approaches that combine non-destructive three-dimensional imaging with quantitative crystallographic analysis. For measuring crystalline properties and orientations, X-ray diffraction techniques are unmatched due to their ability to probe the interior of the sample. In particular, scanning probe tomography approaches with highly focused synchrotron radiation are able to image a large field of view with a 3D spatial resolution down to several hundreds of nm. These techniques have been recently deployed to study for example the properties of bone (5, 26–29) or the osseointegration process around implants (30, 31).

Here, we use X-ray texture tomography combined with X-ray diffraction computed tomography and Rietveld refinement to resolve crystalline texture and structural parameters in 3D across multiple silica-witherite biomorph morphologies. This approach reveals pronounced spatial variations in crystallite orientation, size, shape, and lattice parameters at the nanoscale within individual structures. By comparing different morphologies, we identify recurring crystalline regimes and orientation motifs that are shared across some but are distinct in others. These measurements provide a structural framework for biomorph organization and a basis for evaluating future mechanistic models of morphogenesis.

**Results**

Silica-witherite biomorphs with different morphologies (cf. Fig 1) were synthesized following established procedures (32, 33), sampling a broad morphology space (Fig S1, Fig S2). Representative structures were analyzed by X-ray diffraction to probe the $BaCO_3$ nanocrystallites within the biomorphs (TEM-image of nanocrystallites in Fig 1e). Smaller structures (leaf- and coral-shaped sample; Fig 1a,b), were fully imaged, whereas larger ones (double- and single-helix; Fig 1c,d), were examined in representative regions.

To resolve crystallographic orientation within these structures, we employed texture tomography, which reconstructs spatially resolved orientation distribution functions (ODFs) from two-dimensional diffraction patterns (Fig 2a,b) (34). ODFs describe the statistical distribution of crystallite orientations within a given corresponding subvolume (voxel). In our samples, voxels (Fig 2c) are cubes of 250 to 500 nm side length, therefore representing about 100 to 1000 crystallites.

Across all samples, analysis of the ODFs reveals a preferred orientation of crystallites along their crystallographic *c*-axis and rotational lack of orientation around this axis, (Fig S3), consistent with a fiber texture. Such an arrangement closely resembles a nematic phase. We therefore describe the texture using the nematic director $\vec{n}$ and the nematic order parameter $S$(35). $\vec{n}$ marks the mean direction of the particles' long axes that coincide with the crystallites *c*-axis, while $S$ measures the degree of alignment of these axes, ranging from 0 (no directional order) to 1 (fully aligned). Note that these parameters are insensitive to the orientations of the crystallographic *a*- and *b*-axes.

In addition to orientation information, crystallite size, shape, and lattice parameters were determined for each voxel by Rietveld refinement (36) of reconstructed one-dimensional diffraction patterns (37). This provides access to unit-cell parameters as well as effective crystallite size and size anisotropy. The extracted crystallite volumes and shapes therefore reflect ensemble-averaged properties within each voxel. Across all samples, unit-cell volumes range from 306 $Å^3$ to 309 $Å^3$, slightly exceeding the literature value reported for geological witherite (304 $Å^3$) (38, 15). Crystallite shape was approximated using a biaxial ellipsoid model (Fig 2c, bottom panel), yielding crystallite volumes $V_{crystallite}$ between $1.2 \cdot 10^4$ and $2.0 \cdot 10^4$ $nm^3$ and anisotropies $A_{crystallite}$ from 1.1 to 2.1. These values correspond to apparent crystallite sizes of 20–40 nm and diameters of 10–20 nm. Because X-ray diffraction measures the size of coherently scattering domains, this size is smaller than the particle size, likely containing multiple domains as obtained from imaging-based techniques such as transmission electron microscopy. In fully imaged structures (double helix, coral-like, and leaf-like morphologies), regions of elevated crystallite density and distinct crystalline properties were interpreted as nucleation centers. Spatial variations in orientational ($\vec{n}$, $S$) and structural ($V_{uc}$, $V_{crystallite}$, $A_{crystallite}$) parameters were then analyzed relative to these reference points for radially growing morphologies and with respect to a central axis for helical structures. The analysis is described in more detail in the Supplementary Information (Fig. S4).

*Leaf*

The leaf-shaped structure (Fig 3a) consists mostly of a thin sheet extending radially from a nucleation region at the tip (domain 0) over roughly 50 µm. In addition to the main sheet, a second, shorter sheet emerges from the same nucleation zone and curls after about 10 µm (marked by S). In the following, the direction defined by a vector from this point to a given location in the sheet is referred to as growth direction. Directions orthogonal to the growth direction are labelled transverse (parallel to the sheet) and normal (with respect to the sheet). In contrast to frequently reported flat sheets (2, 23), the sheet presented here is bent normal to the main growth direction, best observed in the transversal views (T) in Fig 3. The complete results of the Rietveld refinement are presented in Fig S5.

The spatial organization of crystallites, visualized via flow-lines of the nematic director (Fig 3b), shows that from the nucleation center (0) towards the top (I, center of II), crystals are aligned with the growth direction. To both sides, crystal orientations gradually deviate while crystallite density $\rho$ decreases (II, III). To the right, the angular spread of the sheet stops where the crystals are aligned in transverse direction (IV). To the left of II, we observe a discontinuity of the growth direction, which aligns with the kink. In domain S crystals are aligned in normal direction and the crystalline density is comparatively high, as in the nucleation center.

A longitudinal slice, i.e. along the growth direction (Fig 3b, L), reveals that crystallite orientations remain largely parallel even as the sheet bends (domain II). Rather than following the macroscopic curvature, the nematic director field maintains a staggered but coherent alignment. In the second sheet (S), the crystallites likewise retain the initial alignment close to the nucleation zone, even as the sheet begins to curl.

These orientational variations are accompanied by systematic changes in crystalline properties (Fig 3c-f). Crystallites at and near the nucleation center are large, highly ordered, and anisotropic, with reduced unit cell volume compared to the rest of the volume. Along the main growth direction (domain I-II), alignment remains high, whereas crystallite size and anisotropy decrease while the unit cell volume increases slightly. At intermediate distances, reduced alignment coincides with an increase in anisotropy and a broader size distribution (domain III). Toward the edges of the structure (domain IV), crystallite size becomes more uniform and anisotropy decreases further. In domain S crystallites align normal to the sheet, the order parameter exhibits oscillatory behavior, and localized regions of large crystallites with reduced anisotropy and increased unit cell volume are observed. Overall, the leaf morphology exhibits a nucleation-centered organization with gradual spatial evolution of both orientation and structural parameters.

*Coral*

Coral-shaped morphologies originate from a central nucleation region and grow roughly isotropically, while developing intricate patterns normal to the growth direction (39). Consistent with the adopted synthesis protocol by Kellermeier et al. (33), the investigated sample (Fig 4) consists of sheet- and cone-like segments without the often reported stem-like features (2, 20, 33, 39). Further, the outer edges are covered with crystallites that likely result from secondary crystallization processes which can occur after the formation of a silica-witherite biomorph is completed (16, 33).

The bottom view of the structure (Fig 4b) shows a central region (domain 0, marked by a dotted outline) of elevated crystallite density, consistent with the nucleation center. From this region outwards, three distinct kinds of domains can be identified. Leaf-like domains (L) are mainly found in the bottom part of the structure, show an initial alignment along the growth direction, followed by a gradual transition to transverse orientations. Cone-like regions (C) are dominated by transverse orientations with a curved director field bending around a central axis. Branched regions (D), similar to L-domains, exhibit alignment along the growth directions with progressive tilting toward the periphery, however in the opposite sense. This coexistence of different motifs within a single structure reflects the highly heterogeneous orientational landscape within the coral-like structure.

Consistent with this heterogeneity, the nematic order parameter remains low across most of the sample, indicating weak global alignment of crystallites. Variations in crystalline properties reveal a core–shell–like organization. The nucleation center (0) contains relatively large, anisotropic crystallites with reduced unit cell volume, closely resembling the nucleation center of the leaf-like morphology. At intermediate distances, both particle size and anisotropy decrease, while the unit cell volume increases. While the L, D, and C subdomains differ in their orientation fields, they show only small differences in crystalline parameters. The dominant contrast is instead between the dense center, the intermediate shell, and the outer edge, where the latter is likely due to secondary crystallization. These variations are spatially heterogeneous and do not exhibit a clear directional pattern. At the structures' edge, crystalline properties resemble those of the central region. This supports the assumption of the presence of crystals resulting from secondary crystallization, although the crystallite density is significantly lower. Overall, the coral morphology is highly heterogeneous, as reflected in the broad range of measured parameters. A more in-depth presentation of the Rietveld refinement results can be found in Fig S6.

### *Double-helix*

The sample presented in Fig 5 consists of a double-helix, and a hollow region, which are both attached to a dense ellipsoidal structure between them (Fig 5a). This region exhibits the highest crystallite density (Fig 5b) and is therefore identified as the nucleation site (domain 0). From this, the double helix extends outwards with a pitch and diameter of around 20 µm, which gradually tightens to about 10 µm in diameter and a pitch around 5 µm. The hollow region extends in the opposite direction and is bounded by thin walls that close toward its termination. The upper part of the structure shows crystallites oriented largely away from the nucleation center, whereas much of the helical strand itself is dominated by transverse orientations. The full Rietveld refinement results can be found in Fig S7.

The crystallites in the double helix are predominantly oriented transverse to the helix axis (Fig 5b), except for the nucleation center, the hollow region as well as the center of the helix at the end of the structure (best visible in supporting movie S3, at 6 s). The nematic order parameter reaches values up to 0.6 in highly ordered regions, particularly within domain 0 and between the double-helix strands (Fig 5c). The highest values of the nematic order parameter are not confined to domain 0, but also occur in the walls of the upper part of the structure and in the region between the two strands. These roughly coincide with large crystallites and low anisotropy. The unit cell volume exhibits moderate variation, ranging from 306 Å$^3$ to 307.5 Å$^3$ (Fig 5f). Due to the intricate structure of this sample, a more detailed comparison to the single helix is given below.

### *Single-helix*

The measured single-helix structure (Fig 6) represents a fragment of a larger morphology and does not include the nucleation center. It is characterized by a helical arrangement extending along a central axis where material is most likely deposited outward in a helical manner, as previously reported(40). This results in a rolled-up structure with a diameter of 20 µm and a pitch of ~5 µm (Fig 6a).

Within this fragment, crystallite orientation gradually transitions from alignment along the central axis (Fig 6a, white dash-dotted line) near the core to predominantly transverse orientations at larger radial distances (Fig 6b). Crystallite density is relatively uniform throughout the structure, decreasing only near the surface. The nematic order parameter is highest near the central axis and oscillates along the single-helix (Fig 6c). Particle size follows a similar but not identical trend, with smaller crystallites near the central axis (Fig 6d), although this region does not fully coincide with the most highly ordered domains. Crystallite anisotropy and unit cell volume display a more complex alternating pattern in the central region (Fig 6e,f), which we will discuss further below. Full Rietveld refinement results are shown in Fig S8.

### *Comparison of single- and double-helix*

A comparison of the double- and single-helix morphologies (Fig 7) reveals both commonalities and differences in their crystallographic organization. In both cases, crystallites transition from alignment along the central axis near the core to transverse orientations with increasing radial distance.

Despite morphological differences, both structures exhibit similar ranges of order parameter and crystallite size. However, approximately isotropic crystallites are observed only in the double helix, and variations in unit cell volume are more pronounced. Based on spatial trends, five regions can be distinguished across both morphologies. Region I is located close to the central axis of the helices, where crystallites are aligned along the axis and highly ordered. In this region, crystallites are large, show low anisotropy and have a reduced unit cell volume. Region II is adjacent to Region I and characterized by a transition in crystalline parameters. In Region III, crystallites are oriented transverse to the helix axis and display a significantly lower degree of order compared to Regions I and II. In Region IV, the crystalline properties start to resemble those of Region I, while the orientation remains similar to Region III. Region V, observed in the single-helix and at the nucleation of the double-helix, is characterized by large and moderately anisotropic crystallites. Regions I-IV recur in a spatially ordered sequence across both helices, with neighboring regions either sharing similar texture or crystalline parameters, resembling a cyclic progression.

These observations indicate that, despite differences in global morphology and the absence of a visible nucleation center in the single helix fragment, both systems share a closely related spatial organization of crystallographic parameters, suggesting a common underlying growth and reorientation mechanism.

Overall, the results indicate that leaf-like and helical morphologies preserve stronger structural coherence around a central aligned region, whereas the coral-like morphology is more fragmented and heterogeneous.

**Discussion**

The present results describe the local crystalline properties across the four main silica–witherite biomorph morphologies. Although the measurements were conducted on fully developed structures, the three-dimensional resolution of the data allows for partial inference of their growth-related organization. Each morphology exhibits a distinct crystallographic signature, while several common structural features emerge across morphologies. A schematic summary of the main findings is provided in Fig. 8.

*Growth motifs*

Across all morphologies, a recurring structural motif is the presence of regions with elevated crystallite alignment that can be associated with central axes or localized regions of structural organization. In leaf-like and helical morphologies, these regions act as structural reference axes from which orientations progressively diverge. In coral morphologies, several regions with such alignment are present locally but do not extend into a globally consistent orientation field.

In the leaf-like structure, this region is located at the tip from which the structure expands. The nematic director field indicates a dominant orientation axis that passes through the nucleation center, with crystallite orientations gradually fanning out laterally. Although the exact local growth trajectory cannot be reconstructed from the fully developed structure alone, the observed stacking direction (red arrow) of crystallites is broadly consistent with outward growth from the tip. At the edges, the directors become transverse to the edge, coinciding with regions where growth appears either to terminate or to transition into alternative local motifs, such as staggered and locally curved domains with related crystallographic organization.

The coral-like structure exhibits a wide range of local structural motifs, including leaf-like, cone-like, and branched domains. Leaf-like and cone-like motifs qualitatively resemble features observed in leaf morphology, namely the main sheet structure and the staggered, locally curved domain. Despite these similarities, the coral structure does not exhibit the same degree of spatial organization. Instead, the coexistence of multiple motifs with weakly correlated orientations indicates a more heterogeneous organization, in which no single dominant growth direction governs the overall structure.

In both helices, the stacking direction is oriented perpendicular to the cross-section of the strands. In the double helix, this results overall in an alignment approximately parallel to the helix axis. In contrast, the single helix with its snail-shell-like architecture, the stacking direction remains predominantly perpendicular to the helix axis. As in the leaf structure, a highly textured central region is present, in both cases oriented along the helix axis, from which the nematic directors progressively diverge towards the outer regions. The observed patterns indicate radially or axially organized growth motifs, where local orientation fields evolve continuously across the structure.

In leaf and helical morphologies, this results in a high degree of structural coherence, whereas in the coral-like morphology, the organization is more fragmented.

### *Crystallographic properties*

Spatial variations in crystallite size, anisotropy, and unit cell parameters are consistently observed across morphologies and often correlate with local orientation and position, but are also characteristic for most morphologies (Fig S9). In all morphologies, regions associated with central alignment tend to contain larger, more elongated crystallites with reduced unit cell volumes. Moving away from these regions, crystallite size decreases, and anisotropy is reduced, although the spatial organization of these variations differs between morphologies.

Leaf-like structures exhibit a more pronounced spatial stratification of crystallographic properties, whereas coral morphologies show greater variability without clear spatial partitioning. Helical morphologies display systematic transitions in crystallographic parameters along radial directions and along the helix axis.

These observations indicate that crystallite growth is coupled to local structural organization and that variations in crystallographic properties reflect differences in growth conditions across the structure.

### *Structural implications for growth*

The results indicate that the distinct morphologies share common local structural motifs, and that their global architecture may reflect differences in how these motifs are spatially organized across length scales. This is consistent with concepts from non-classical crystallization and biomineralization literature.(41, 42)

The radial or axial organization of crystallite orientations, together with the presence of central aligned regions, is consistent with structural features reported for spherulitic-like growth, in which building units assemble into hierarchically organized structures with spatially varying orientation fields.(43) Here, this comparison serves as a structural analogy and does not imply a specific growth mechanism.

Leaf-like morphologies are generally considered to grow on surfaces, which naturally favors a planar morphology. The measured structure displays a central aligned region within a flat domain, consistent with growth under geometrical confinement. The relatively limited spatial extent of these structures suggests that sustained growth depends on maintaining effective coupling between neighboring structurally distinct regions, with reductions in such coupling potentially leading to growth termination or transitions to alternative morphologies.

Helical morphologies exhibit continuous internal orientation along curved paths, indicating that local crystallographic order can persist even in strongly curved structures. Single and double helices thus appear as closely related variants of the same general structural organization. In the sample studied here, the double helix starts from an initial aggregate with an apparent twofold symmetry, from which two strands emerge. In contrast, single helices may form when only one of these growth directions is maintained, for example because of local boundary conditions or perturbations. At the same time, this is unlikely to be the only route to helix formation, as helical morphologies have also been reported to transform into one another or to arise from other biomorph shapes.

The coral-like structure shows greater structural heterogeneity, with multiple domains lacking a dominant orientation axis. Furthermore, the relatively chaotic arrangement of domains likely does not favor such coupling between neighboring regions, resulting in fragmented orientation fields and diminished long-range coherence.

In addition to these structural considerations, prior work suggests that the silicate phase contributes to shaping the local growth environment. Silicate species can form polymerized networks whose properties depend on reaction conditions, influencing transport, confinement, and interfacial interactions.(44) Interactions between silicate and carbonate surfaces have been proposed to promote oriented nucleation and interfacial coupling, potentially contributing to correlated crystallographic orientations (15). Within this framework, the silicate matrix modulates spatial and kinetic aspects of crystallization rather than acting as an inert scaffold (16, 21). Morphology selection is likely influenced by the interplay between carbonate supersaturation and silicate condensation(16). Supersaturation controls nucleation and growth kinetics, while silicate condensation affects matrix structure, including permeability and interfacial characteristics. Variations in the relative rates of these processes may therefore impact nucleation density, domain coupling, and the degree of structural heterogeneity. This sensitivity to environmental conditions is consistent with prior studies showing that variations in reaction–diffusion dynamics and supersaturation lead to distinct hierarchical architectures in carbonate–silicate systems (2, 16, 22), although these processes are not directly resolved in the present work.

Overall, the results indicate that morphology is governed by the spatial organization and coupling of local crystallographic motifs within an evolving silicate-modulated growth environment. Differences between morphologies can be understood in terms of variations in structural confinement, boundary conditions, and the balance between carbonate precipitation and silicate network formation. While these observations provide structural insight, further in situ investigations will be required to directly resolve growth dynamics and clarify how specific environmental parameters govern morphology selection.

## Materials and Methods

### *Silica Biomorph synthesis:*

For the silica–witherite biomorph synthesis, following the procedure described by Kellermeier *et al.*(32), equal volumes of a dissolved silica solution (16.8 mM) and 10 mM barium chloride solution (prepared from barium chloride dihydrate purchased from Sigma-Aldrich, ≥99%) were prepared using Milli-Q water (18 MΩ·cm). For this, the dissolved silica solution was freshly diluted 1:350 (v/v) from a concentrated water glass stock solution (Sigma-Aldrich, reagent grade, 27% as $SiO_2$) and its pH was adjusted to 11.3 using small aliquots of 0.1 M NaOH (Merck). Crystallization experiments were carried out in small open vessels—typically 24-well plates—by rapidly mixing 1 mL of the silicate solution with 1 mL of the $BaCl_2$ solution. The initial pH of the mixture was adjusted to 11 using 0.1 M hydrochloric acid (Merck), placing the system in the biomorph-forming regime. Immediately after mixing, wells were loosely covered with a lid and kept undisturbed at room temperature (≈25 °C). This setup allows atmospheric $CO_2$ to diffuse slowly into the solution over several hours, generating carbonate in situ and gradually lowering the pH. After 8-10 h, the solutions are removed. The formed structures are then washed several times with first water then ethanol before they are dried under air. Under these conditions, a mixture of morphologies, including leaf-like sheets and single and double helices is obtained.

Coral-like silica-witherite biomorphs were synthesized by adopting another protocol reported by Kellermeier et al.(33) In this case, the concentration of the barium chloride solution was increased to 500 mM and the initial pH after mixing with the dissolved silica solution was lowered to 10.5. The rest of the procedure is analogous to the one described above.

For texture-tomography measurements, representative morphologies were selected under a stereomicroscope, isolated manually using an eyelash-based manual manipulator, and mounted individually on glass needles with a minute amount of Araldite Standard epoxy and oriented with the help of said eyelash manipulator. This ensured stable mechanical support while preserving the native crystalline texture of the delicate biomorph structures.

*Texture tomography data acquisition:*

X-ray diffraction experiments were carried out at the ESRF-EBS beamline ID13 EH3 nanobranch. A 15 keV (wavelength $\lambda$ = 0.0826 nm) X-ray beam was produced by a channel-cut Si(111) monochromator and pre-focused by a set of compound beryllium lenses onto the final focusing optics, a set of multi-layer Laue lenses (45), producing a beam of ~300 x 300 nm with a photon flux of $10^{12}$ photons per second at the sample position. The sample was mounted on a custom designed goniometer(28, 29) based on SmarAct actuators and scanned by a piezo stage. The diffraction signal was recorded with an Eiger X 4M and with a sample-to-detector distance of ~150 mm. The primary beam was blocked by a ~300 µm lead beamstop. The setup gave access to the usable $q$-range 0.5–32 nm$^{-1}$ that at detector corners extended up to 40 nm$^{-1}$. The momentum transfer vector $q$ is defined as $q=\frac{4\pi sin\,(\theta)}{\lambda}$ where $\lambda$ is the x-ray wavelength and $\theta$ the scattering angle. For each projection, the samples were scanned with a step size between 250 and 500 nm in a continuous scanning mode and an adaptive field of view that enabled us to catch the full sample while avoiding excessive air regions around the sample to be scanned. The full experimental details for each sample are provided in Table 1. Diffraction patterns were recorded with 2 ms exposure time. Subsequently, the sample was rotated around the $z$ axis and tilted around the $y$ axis. A total of ~260 projections (details in Table 1) were collected for 10 tilt angles $\varkappa$ between 0° and 45°. At $\varkappa$ = 0°, rotation angles were collected at equal distances between 0 and 180°, for $\varkappa$ > 0° between 0 and 360°. The number of rotation angles was reduced by a factor of cos $\varkappa$ for the respective tilt.

The X-ray dose $d$ deposited in each sample was calculated according to(46) as:

$$d=\frac{\mu N_0 \varepsilon}{\rho}$$

where $\mu$ = 269.58 cm$^{-1}$ is the linear absorption coefficient for $BaCO_3$ at 15.2 keV, $N_0$ = $7.07\cdot10^{18}$ photons m$^{-2}$ is the integrated flux per unit area, $\varepsilon$ =$2.44\cdot10^{-15}$ J is the photon energy and $\rho$ = 4.3 g cm$^{-3}$ is the mass density. The dose imparted on each sample is given in Tab. 1.

*Texture tomography data treatment:*

The raw data was regrouped into 100 radial and 120 azimuthal bins over a $q$-range of 0.4 to 35 nm$^{-1}$ using pyFAI(47). To correct for small deviations from the center of the field of view, we averaged each diffraction pattern azimuthally and over the 16 nm$^{-1}$ to 17.5 nm$^{-1}$ and re-aligned the projections using an optical-flow alignment strategy(37, 48). Texture tomograms were reconstructed with the TexTOM software(34), using a hyperspherical harmonic model up to order 8 to describe ODFs and a projected gradient-descend algorithm for parameter optimization.

To extract order parameters, we calculated the ODF in each voxel and extracted the preferred direction of the *c*-axis from the maximum of the distribution. Order parameters and nematic directors were calculated as the largest eigenvalues and the corresponding eigenvector of the order tensor *Q*(35):

$$Q= \sum_i ODF(g_i)\left(\frac{3}{2}\hat{c}(g_i)\otimes\hat{c}(g_i)\text{-}\frac{1}{2}I\right),$$

where $g_i$ is a crystal orientation, $\hat{c}$ the respective unit direction vector of the crystalline c-axis, the *I* is the identity matrix and $\otimes$ is the outer product.

*Rietveld refinement data treatment:*

The XRD-CT position resolved diffractograms were Rietveld refined individually using the MultiRef (36) software package, providing a Matlab interface to GSAS(49) that allows to set up automatic refinements. MultiRef runs several refinement cycles and updates the model automatically according to user settings. An initial Rietveld model was developed using a representative data point of the sample. To take the effects of instrumental broadening into account, the diffraction pattern of the $Al_2O_3$ calibrant (NIST 674a) was Rietveld refined under the assumption that the calibrant produces negligible line broadening and all broadening was modeled using the Gaussian term of the instrumental profile. We want to point out that NIST 674a is an intensity and not a line profile standard, but all available line standards use large crystals which are inherently unsuitable for nanobeam diffraction tomography. The values presented here for line broadening are thus strictly experimental and can be regarded as an upper bound. A W-parameter of 5.03 and 6.9 was refined for the two experimental sessions and used for the respective refinements. In practice, this implies a rather small, almost negligible instrumental broadening compared to the investigated crystallite sizes in the order of 1 nm, which was corrected for nonetheless.

The Rietveld model started from a witherite structure and over the course of 4 refinement steps the complexity of the model was increased.

In the first step, only the scale factor and a $6^{th}$ order Chebyshev polynomial background were fitted. In the second step, a preferred orientation model using spherical harmonics up to an order of 6 was fitted. The third step consisted of fitting the crystal lattice parameters. In the fourth step, the profile parameters were added to the model.

The peak profiles were modelled by a Pseudo-Voigt profile function including an isotropic and an anisotropic size broadening parameter describing the anisotropic shape of the witherite crystallites. The shape was approximated by a prolate ellipsoid with rotational symmetry about its major axis, therefore parametrized by its length and diameter. The *c*-axis was taken as the anisotropic axis. Only data points containing diffraction signals were included in the refinement procedure, other points were masked out.

*X-ray nanotomography*

The X-ray nano-tomography was performed at beamline ID16B (51) of the European Synchrotron Radiation Facility (ESRF) using a propagation-based X-ray phase contrast nano-imaging method with a pink beam ($\Delta E/E \sim 10^{-2}$) with an energy of 29.6 keV. Four tomographic acquisitions are performed at four propagation distances (holotomography technique (52)). For each tomographic scan, 2505 projections are acquired over 360° on a PCO edge 4.2 camera with 2048 × 2048 pixels. The effective pixel size was 50 nm and the acquisition time per projection was 10 ms for Leaf and 20ms for single and double helix. The phase retrieval calculation and reconstruction were done using the software PyNX (53) with a Paganin-based calculation using a delta/β of 250 and Nabu (54) for the filtered back projection reconstruction. In the reconstructed volumes, the remaining ring artifacts were removed using a post-processing Matlab algorithm based on image filtering(55).

**Acknowledgments**

We thankfully acknowledge Helmut Cölfen for initiating this collaboration and many fruitful discussions around the topic of biomorph mineralization.

We acknowledge the ESRF, Grenoble, France for supplying beam time for the experiments and the Partnership for Soft Condensed Matter (PSCM) for support during the preparation of the experiment.

The data presented in this publication is available under the following DOI

https://doi.esrf.fr/10.15151/ESRF-DC-2228679620

and the data processing codes are available from

https://gitlab.fresnel.fr/textom/textom

This work is funded by the European Union, European Research Council Horizon Europe (grant No.101041871). Views and opinions expressed are those of the author(s) only and do not necessarily reflect those of the European Union or the European Research Council. Neither the European Union nor the granting authority can be held responsible for them.

B. Maier was supported by the Deutsche Forschungsgemeinschaft (DFG, project number 14951021). We thank the electron microscopy centre, namely Michael Laumann and Paavo Bergmann, of the biology department at the University of Konstanz for the preparation of the thin-sections and their scientific input. The TEM device used for images of the thin-sections was funded by the DFG (project number 409790724).

HB and AS acknowledge funding from the ESS lighthouse on hard materials in 3D, SOLID, funded by the Danish Agency for Science and Higher Education (grant number 8144-00002B).

**References**

1. C. N. Kaplan, *et al.*, Controlled growth and form of precipitating microsculptures. *Science* **355**, 1395–1399 (2017).

2. W. L. Noorduin, A. Grinthal, L. Mahadevan, J. Aizenberg, Rationally Designed Complex, Hierarchical Microarchitectures. *Science* **340**, 832–837 (2013).

3. A. Menichetti, *et al.*, Local Light-Controlled Generation of Calcium Carbonate and Barium Carbonate Biomorphs via Photochemical Stimulation. *Chemistry A European J* **27**, 12521–12525 (2021).

4. M. H. Bistervels, *et al.*, Light-controlled morphological development of self-organizing bioinspired nanocomposites. *Nanoscale* **16**, 2310–2317 (2024).

5. T. A. Grünewald, M. Liebi, H. Birkedal, Crossing length scales: X-ray approaches to studying the structure of biological materials. *IUCrJ* **11**, 708–722 (2024).

6. J. M. Garcia-Ruiz, On the formation of induced morphology crystal aggregates. *Journal of Crystal Growth* **73**, 251–262 (1985).

7. J. M. García-Ruiz, *et al.*, Self-Assembled Silica-Carbonate Structures and Detection of Ancient Microfossils. *Science* **302**, 1194–1197 (2003).

8. M. Kellermeier, H. Cölfen, J. M. García-Ruiz, Silica Biomorphs: Complex Biomimetic Hybrid Materials from "Sand and Chalk." *Eur J Inorg Chem* **2012**, 5123–5144 (2012).

9. J. Rouillard, J. -M. García-Ruiz, J. Gong, M. A. Van Zuilen, A morphogram for silica-witherite biomorphs and its application to microfossil identification in the early earth rock record. *Geobiology* **16**, 279–296 (2018).

10. P. Cintas, Chasing Synthetic Life: A Tale of Forms, Chemical Fossils, and Biomorphs. *Angew Chem Int Ed* **59**, 7296–7304 (2020).

11. L. Helmbrecht, *et al.*, Directed Emission from Self-Assembled Microhelices. *Adv Funct Materials* **30**, 1908218 (2020).

12. T. Holtus, *et al.*, Shape-preserving transformation of carbonate minerals into lead halide perovskite semiconductors based on ion exchange/insertion reactions. *Nature Chem* **10**, 740–745 (2018).

13. H. C. Hendrikse, *et al.*, Rational Design of Bioinspired Nanocomposites with Tunable Catalytic Activity. *Crystal Growth & Design* **21**, 4299–4304 (2021).

14. J. Opel, *et al.*, Symbiosis of Silica Biomorphs and Magnetite Mesocrystals. *Adv Funct Materials* **29**, 1902047 (2019).

15. E. Bittarello, F. Roberto Massaro, D. Aquilano, The epitaxial role of silica groups in promoting the formation of silica/carbonate biomorphs: A first hypothesis. *Journal of Crystal Growth* **312**, 402–412 (2010).

16. M. Kellermeier, *et al.*, Growth Behavior and Kinetics of Self-Assembled Silica–Carbonate Biomorphs. *Chemistry A European J* **18**, 2272–2282 (2012).

17. J. M. García-Ruiz, E. Melero-García, S. T. Hyde, Morphogenesis of Self-Assembled Nanocrystalline Materials of Barium Carbonate and Silica. *Science* **323**, 362–365 (2009).

18. P. Knoll, O. Steinbock, Nanodot-to-Rod Transition and Particle Attachment in Self-Organized Polycrystalline Aggregates. *Crystal Growth & Design* **19**, 4218–4223 (2019).

19. S. T. Hyde, A. M. Carnerup, A.-K. Larsson, A. G. Christy, J. M. García-Ruiz, Self-assembly of carbonate-silica colloids: between living and non-living form. *Physica A: Statistical Mechanics and its Applications* **339**, 24–33 (2004).

20. J. Opel, *et al.*, Structural Transition of Inorganic Silica–Carbonate Composites Towards Curved Lifelike Morphologies. *Minerals* **8**, 75 (2018).

21. J. Eiblmeier, *et al.*, New insights into the early stages of silica-controlled barium carbonate crystallisation. *Nanoscale* **6**, 14939–14949 (2014).

22. J. Eiblmeier, M. Kellermeier, D. Rengstl, J. Manuel García-Ruiz, W. Kunz, Effect of bulk pH and supersaturation on the growth behavior of silica biomorphs in alkaline solutions. *CrystEngComm* **15**, 43–53 (2013).

23. P. Knoll, E. Nakouzi, O. Steinbock, Mesoscopic Reaction–Diffusion Fronts Control Biomorph Growth. *J. Phys. Chem. C* **121**, 26133–26138 (2017).

24. H. P. Rothbaum, A. G. Rohde, Kinetics of silica polymerization and deposition from dilute solutions between 5 and 180°C in *Journal of Colloid and Interface Science*, (1979), pp. 533–559.

25. O. Weres, A. Yee, L. Tsao, Kinetics of silica polymerization. *Journal of Colloid and Interface Science* **84**, 379–402 (1981).

26. N. K. Wittig, *et al.*, Bone Biomineral Properties Vary across Human Osteonal Bone. *ACS Nano* **13**, 12949–12956 (2019).

27. J. Palle, *et al.*, Nanobeam X-ray fluorescence and diffraction computed tomography on human bone with a resolution better than 120 nm. *Journal of Structural Biology* **212**, 107631 (2020).

28. T. A. Grünewald, *et al.*, Mapping the 3D orientation of nanocrystals and nanostructures in human bone: Indications of novel structural features. *Sci. Adv.* **6**, eaba4171 (2020).

29. T. A. Grünewald, *et al.*, Bone mineral properties and 3D orientation of human lamellar bone around cement lines and the Haversian system. *IUCrJ* **10** (2023).

30. M. Liebi, *et al.*, 3D nanoscale analysis of bone healing around degrading Mg implants evaluated by X-ray scattering tensor tomography. *Acta Biomaterialia* **134**, 804–817 (2021).

31. I. Rodriguez-Fernandez, *et al.*, Physical exercise impacts bone remodeling around bio-resorbable magnesium implants. *Acta Biomaterialia* **193**, 623–631 (2025).

32. M. Kellermeier, *et al.*, Additive-induced morphological tuning of self-assembled silica–barium carbonate crystal aggregates. *Journal of Crystal Growth* **311**, 2530–2541 (2009).

33. M. Kellermeier, "Co-mineralization of alkaline-earth carbonates and silica," Universität Regensburg. (2011).

34. M. P. K. Frewein, *et al.*, Texture tomography, a versatile framework to study crystalline texture in 3D. *IUCrJ* **11**, 809–820 (2024).

35. E. G. Virga, *Variational Theories for Liquid Crystals*, 1st Edition (Chapman & Hall/CRC, 1994).

36. S. Frølich, H. Birkedal, MultiRef : software platform for Rietveld refinement of multiple powder diffractograms from *in situ* , scanning or diffraction tomography experiments. *J Appl Crystallogr* **48**, 2019–2025 (2015).

37. L. C. Nielsen, P. Erhart, M. Guizar-Sicairos, M. Liebi, Small-angle scattering tensor tomography algorithm for robust reconstruction of complex textures. [Preprint] (2023). Available at: http://arxiv.org/abs/2305.07750 [Accessed 4 December 2023].

38. De Villiers, Johan PR, Crystal structures of aragonite, strontianite, and witherite. *American Mineralogist: Journal of Earth and Planetary Materials* **56**, 758--767 (1971).

39. A.-K. Malchow, A. Azhand, P. Knoll, H. Engel, O. Steinbock, From nonlinear reaction-diffusion processes to permanent microscale structures. *Chaos: An Interdisciplinary Journal of Nonlinear Science* **29**, 053129 (2019).

40. E. Nakouzi, P. Knoll, K. B. Hendrix, O. Steinbock, Systematic characterization of polycrystalline silica–carbonate helices. *Phys. Chem. Chem. Phys.* **18**, 23044–23052 (2016).

41. H. Cölfen, M. Antonietti, *Mesocrystals and Nonclassical Crystallization*, 1st Ed. (Wiley, 2008).

42. S. Mann, Biomineralization and biomimetic materials chemistry. *J. Mater. Chem.* **5**, 935 (1995).

43. A. G. Shtukenberg, Y. O. Punin, E. Gunn, B. Kahr, Spherulites. *Chem. Rev.* **112**, 1805–1838 (2012).

44. M. Matinfar, J. A. Nychka, A review of sodium silicate solutions: Structure, gelation, and syneresis. *Advances in Colloid and Interface Science* **322**, 103036 (2023).

45. S. Niese, *et al.*, Full-field X-ray microscopy with crossed partial multilayer Laue lenses. *Opt. Express* **22**, 20008 (2014).

46. M. R. Howells, *et al.*, An assessment of the resolution limitation due to radiation-damage in X-ray diffraction microscopy. *Journal of Electron Spectroscopy and Related Phenomena* **170**, 4–12 (2009).

47. J. Kieffer, *et al.*, Application of signal separation to diffraction image compression and serial crystallography. *J Appl Crystallogr* **58**, 138–153 (2025).

48. M. Odstrčil, M. Holler, J. Raabe, M. Guizar-Sicairos, Alignment methods for nanotomography with deep subpixel accuracy. *Opt. Express* **27**, 36637 (2019).

49. A. C. Larson, R. B. Von Dreele, "General Structure Analysis System (GSAS)" (2004).

50. C. M. Holl, J. R. Smyth, H. M. S. Laustsen, S. D. Jacobsen, R. T. Downs, Compression of witherite to 8 GPa and the crystal structure of BaCO 3 II. *Physics and Chemistry of Minerals* **27**, 467–473 (2000).

51. G. Martínez-Criado, *et al.*, ID16B: a hard X-ray nanoprobe beamline at the ESRF for nano-analysis. *J Synchrotron Rad* **23**, 344–352 (2016).

52. P. Cloetens, *et al.*, Holotomography: Quantitative phase tomography with micrometer resolution using hard synchrotron radiation x rays. *Applied Physics Letters* **75**, 2912–2914 (1999).

53. V. Favre-Nicolin, *et al.*, *PyNX* : high-performance computing toolkit for coherent X-ray imaging based on operators. *J Appl Crystallogr* **53**, 1404–1413 (2020).

54. P. Paleo, *et al.*, Nabu 2024.1. (2024). https://doi.org/10.5281/zenodo.11104029. Deposited April 2024.

55. A. Lyckegaard, G. Johnson, P. Tafforeau, Correction of Ring Artifacts in X-ray Tomographic Images. *International Journal of Tomography and Statistics* **18**, 1–9 (2011).

56. S. R. Stock, F. De Carlo, J. D. Almer, High energy X-ray scattering tomography applied to bone. *Journal of Structural Biology* **161**, 144–150 (2008).

57. P. Bleuet, *et al.*, Probing the structure of heterogeneous diluted materials by diffraction tomography. *Nature Materials* **7**, 468–472 (2008).

## Figures and Tables

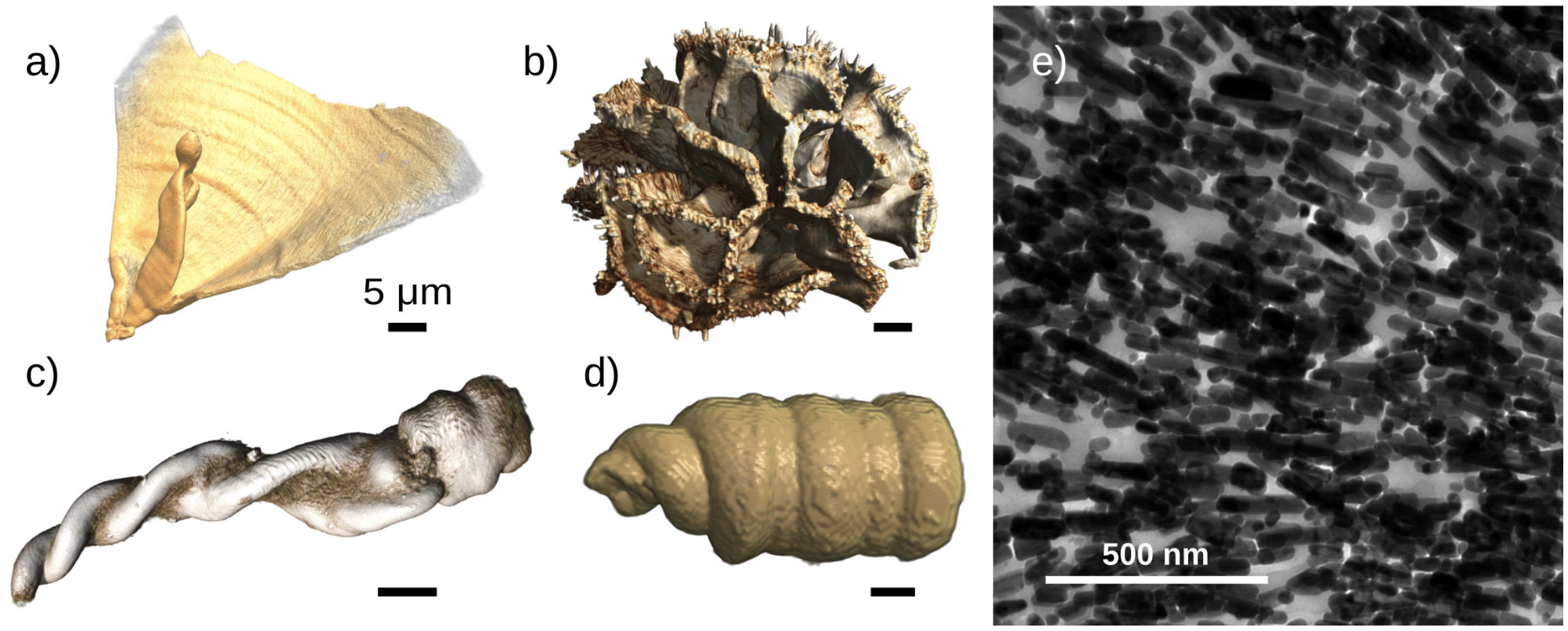

**Figure 1.** Morphology of silica-witherite biomorph structures studied in this work. X-ray nanotomography images of biomorph samples: leaf (a), coral (b), double helix (c). d) Tomogram from texture tomography measurements of the single helix structure (this sample could not be recovered after the experiment). Scale bars are 5 µm. e) TEM image of a slice of a double-helical biomorph, showing the witherite crystals in black with amorphous silica in the interstices.

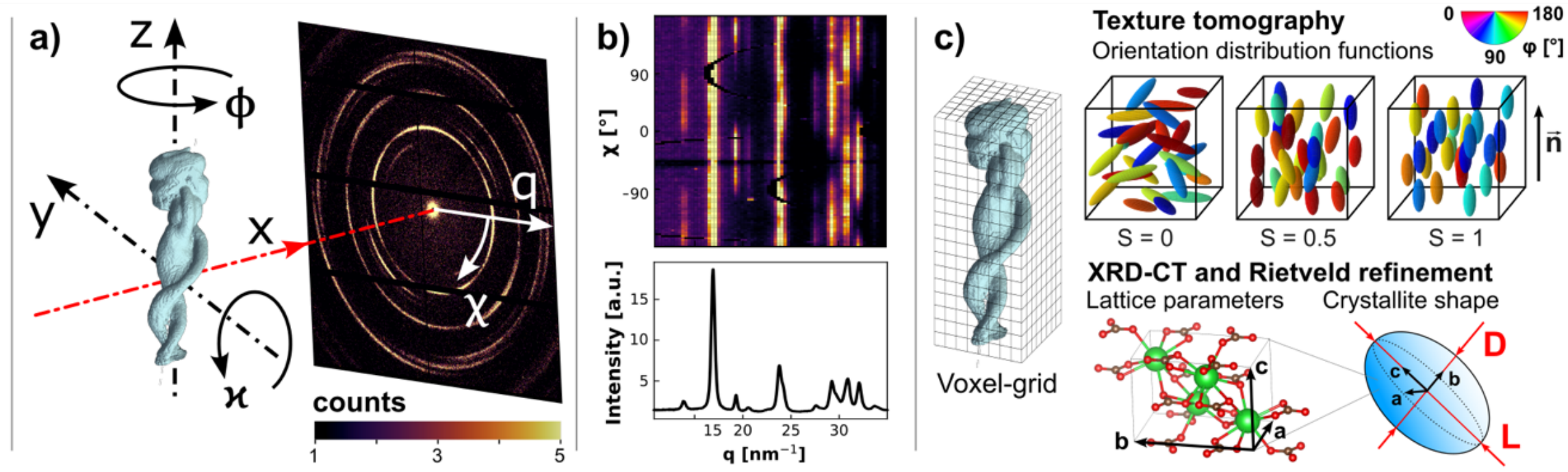

**Figure 2.** Experimental and analysis strategy for texture tomography and XRD-CT/Rietveld refinement: a) The sample is mounted on a 2-axis goniometer (φ, ϰ) with the possibility of horizontal and vertical translations (y, z). It is illuminated by a focused X-ray beam (x-direction) and diffraction patterns are recorded on a 2D-pixel detector. The full biomorph is scanned in a 2D grid by translating the sample for each sample orientation (φ, ϰ). B) The data is radially and azimuthally rebinned and inverted using texture tomography (34) and XRD-CT/Rietveld refinement approaches (56, 57, 36). c) This panel shows how a biomorph is divided into regular sub-volumes (voxels). For each voxel we obtain the local orientation distribution and orientation-averaged crystalline properties. The upper panel shows sketches of possible particle distributions and how they relate to the nematic director $\vec{n}$ (average direction of long crystallite axes) and the order parameter *S* (0 for completely disordered, 1 if all crystallite long axes are fully aligned). The color-coded angle ϕ is the rotation angle around of the [1,0,0] axis around the long axis of the particle. From Rietveld refinement the mean lattice parameters (*a,b,c*) and mean particle length L and diameter *D* are obtained. In the following, we refer to crystalline properties in form of the unit cell volume $V_{UC}=abc$, crystallite volume $V_{Crystallite}=\frac{1}{6}\pi D^2 L$ and crystallite anisotropy $A_{Crystallite}=L/D$.

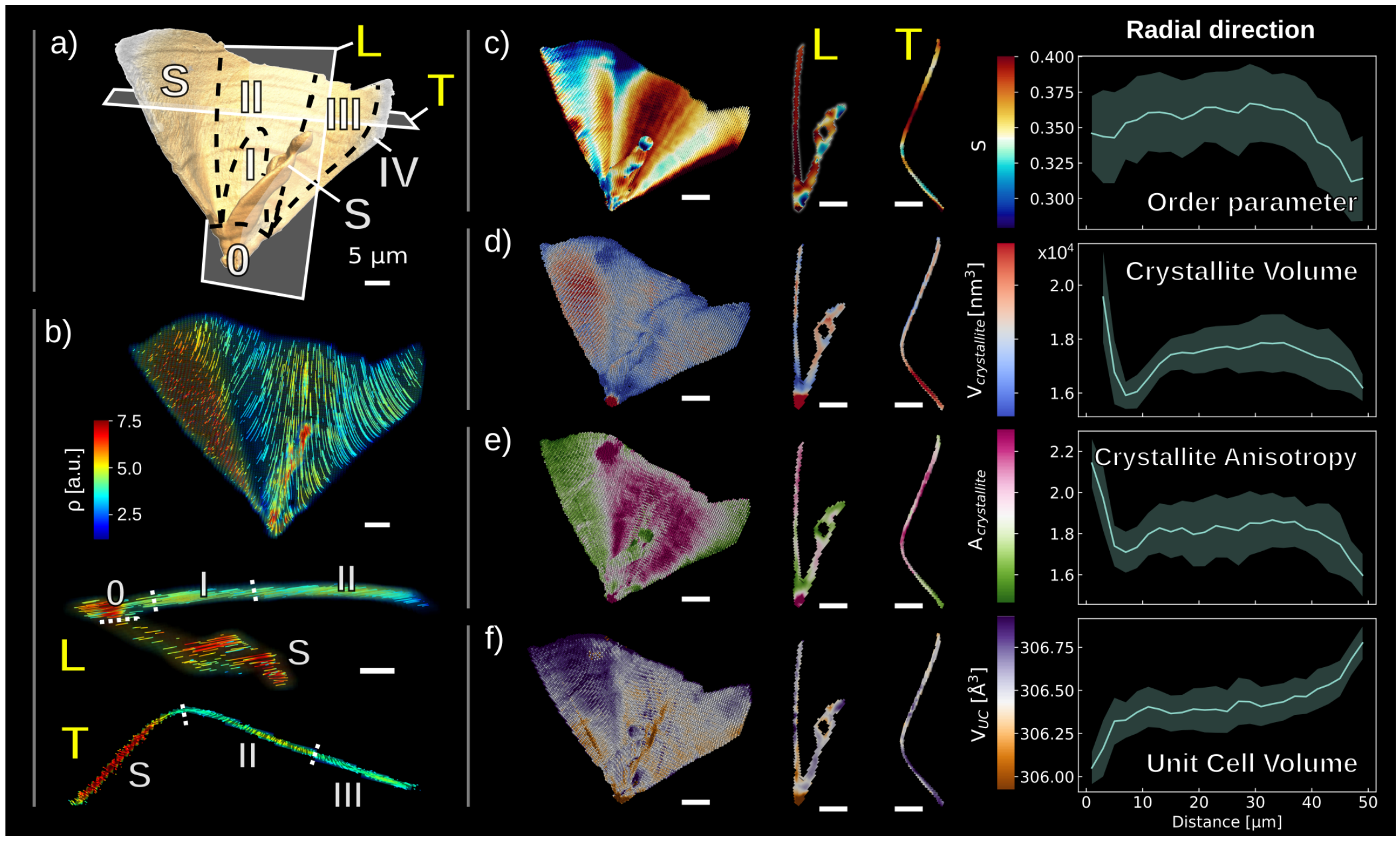

**Figure 3.** Texture tomography and XRD-CT/Rietveld results for the leaf-like morphology. A) Overview of the 3D structure with longitudinal (L) and transversal (T) cutting planes. b) Different views of the biomorph with flow-lines drawn along the nematic director obtained from texture tomography, the color scale corresponding to the density of crystalline material ρ. c-f) Local crystalline properties as well as plots of the average values given as function of the distance from the tip of the biomorph (in region 0). The shaded areas correspond to the standard deviation.

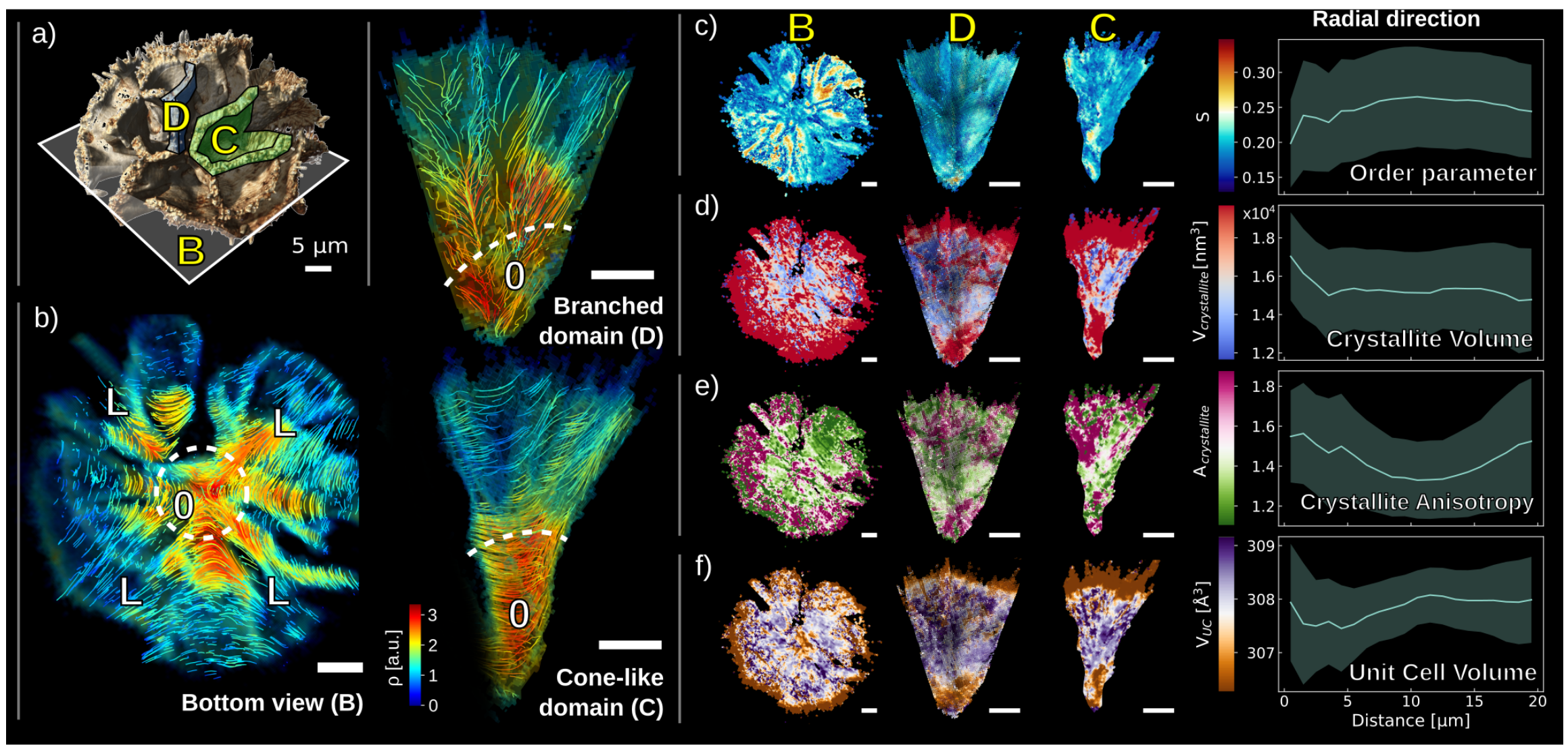

**Figure 4.** Texture tomography and XRD-CT/Rietveld results for the coral-like morphology. a) Overview of the 3D structure with a horizontal cutting plane at the base (B) as well as a branched (D) and a cone-like (C) substructure highlighted. b) Different views of the biomorph with flow-lines drawn along the nematic director obtained from texture tomography, the color scale corresponding to the density of crystalline material ρ. c-f) Local crystalline properties as well as plots of the average values given as function of the distance from the center of the base cutting plane. The shaded areas correspond to the standard deviation.

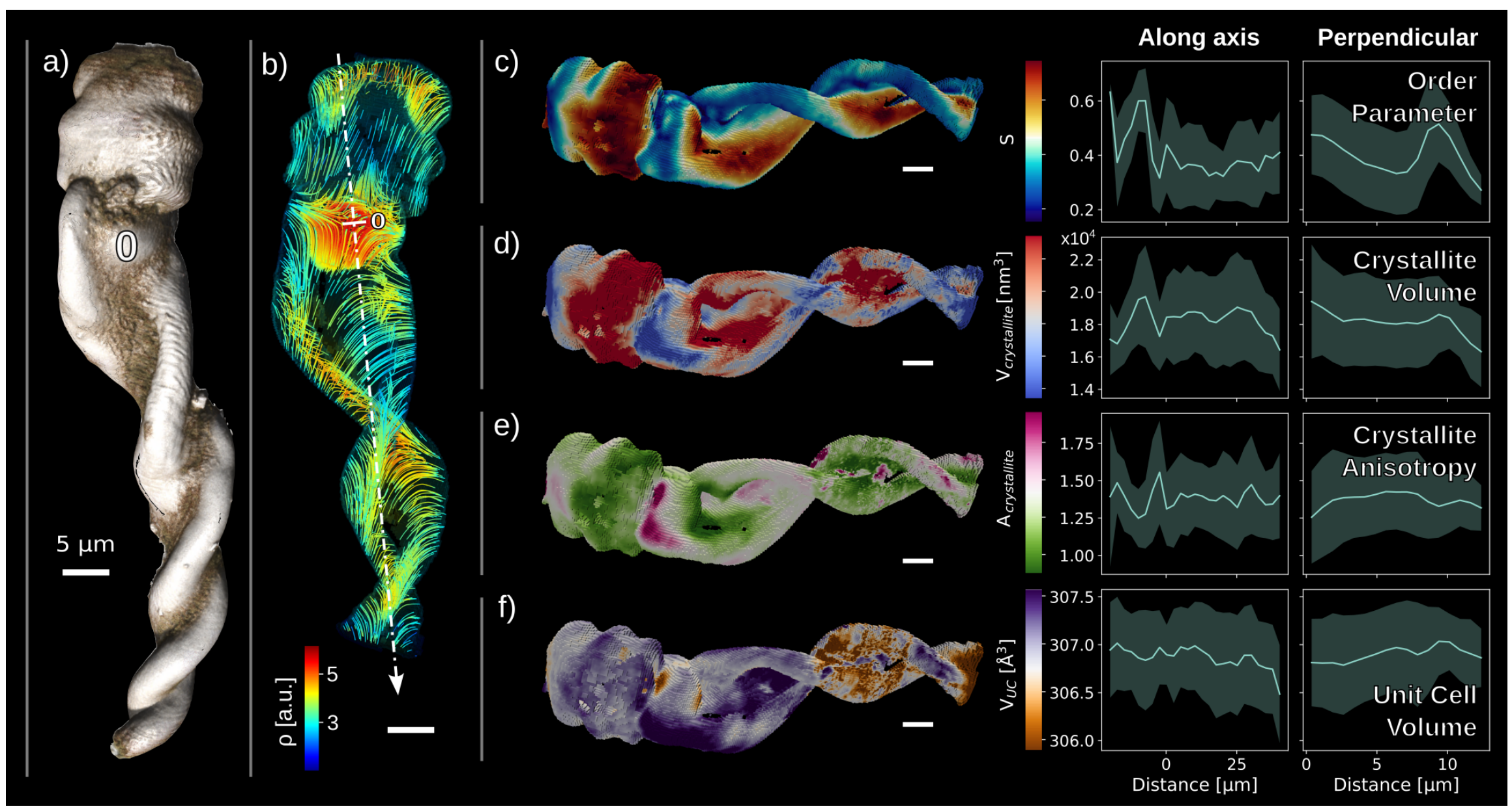

**Figure 5.** Texture tomography and XRD-CT/Rietveld results for the double helix-like morphology. a) Overview of the 3D structure. b) Biomorph with flow-lines drawn along the nematic director obtained from texture tomography, the color scale corresponding to the density of crystalline material ρ. c-f) Local crystalline properties as well as plots of the average values along the central axis shown in b), as well as perpendicular to the axis. The shaded areas correspond to the standard deviation.

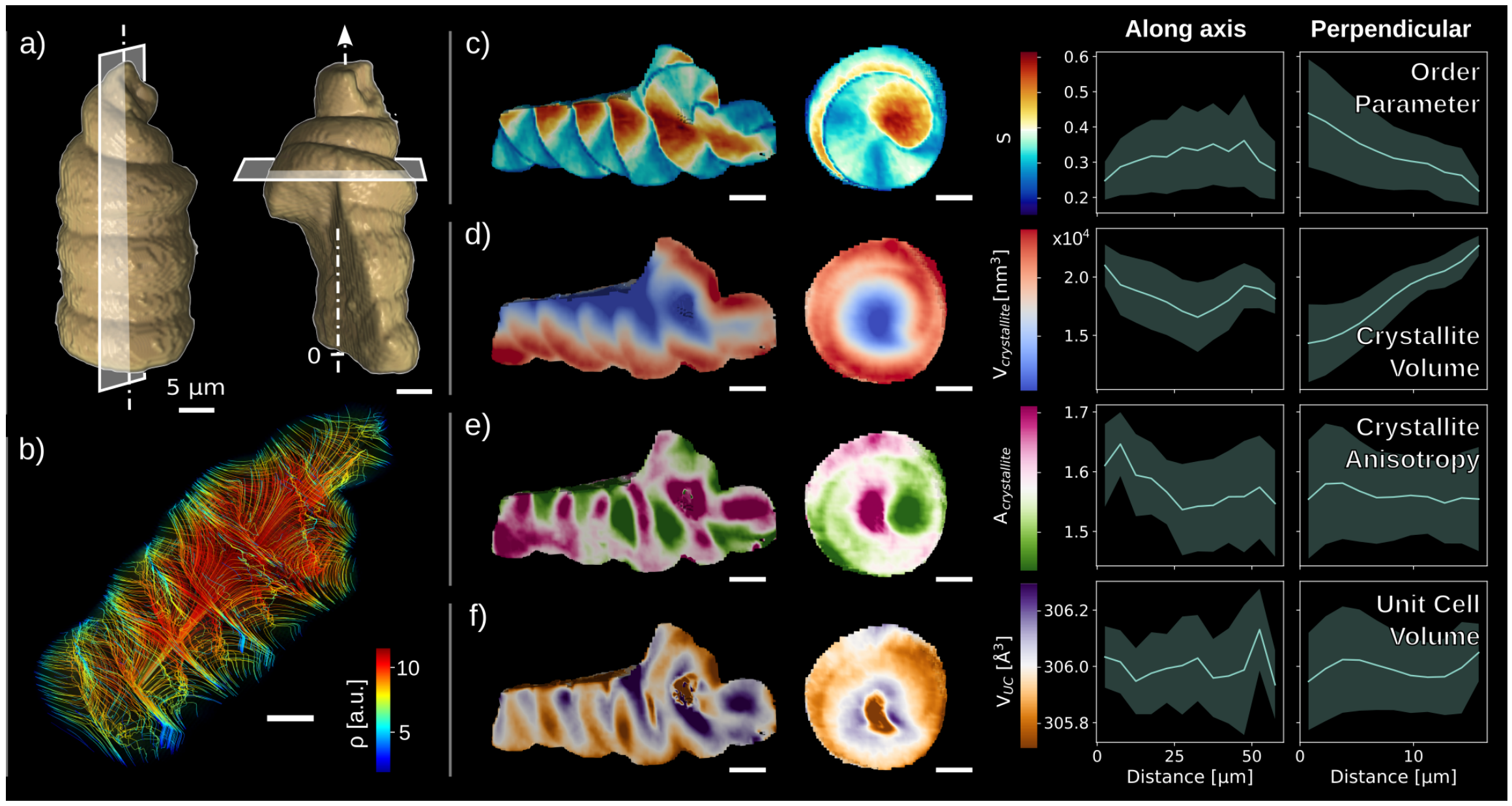

**Figure 6.** Texture tomography and XRD-CT/Rietveld results for the single helix-like morphology. a) Overview of the 3D structure from two angles with the central axis marked by a dotted line and cutting planes used for the representations in subfigures c-f). b) Biomorph with flow-lines drawn along the nematic director obtained from texture tomography, the color scale corresponding to the density of crystalline material ρ. c-f) Local crystalline properties as well as plots of the average values along the central axis shown in a), as well as perpendicular to the axis. The shaded areas correspond to the standard deviation.

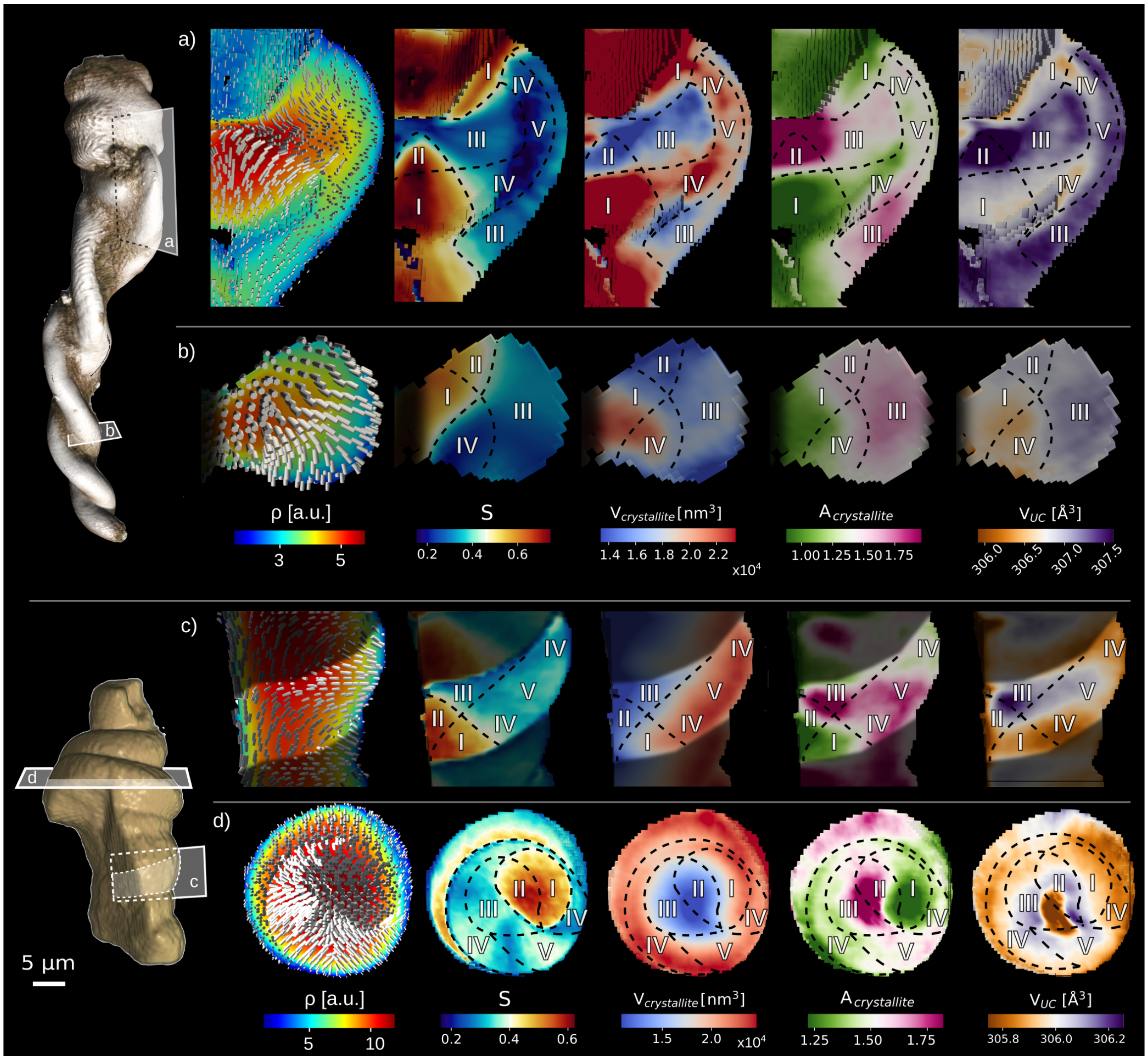

**Figure 7.** Comparison between different parts of the double- and single helix structures. The crystalline properties (crystalline density $\sigma$, order parameter S, crystallite volume $V_{Crystallite}$, crystallite anisotropy $A_{Crystallite}$ and unit cell volume $V_{UC}$) are shown for nucleation zone (a) and the helix strand (b) of the double helix-like structure and a low order region (c) and a helix strand (d) of the single helix-like structure. The overall morphology is shown in the leftmost column with the cross-section planes shown as white rectangles. The [001] orientation of the crystallites are overlaid with the crystallite density white sticks. The orientation and crystalline properties form distinct regions in the sample that could be identified and are labeled with I-V

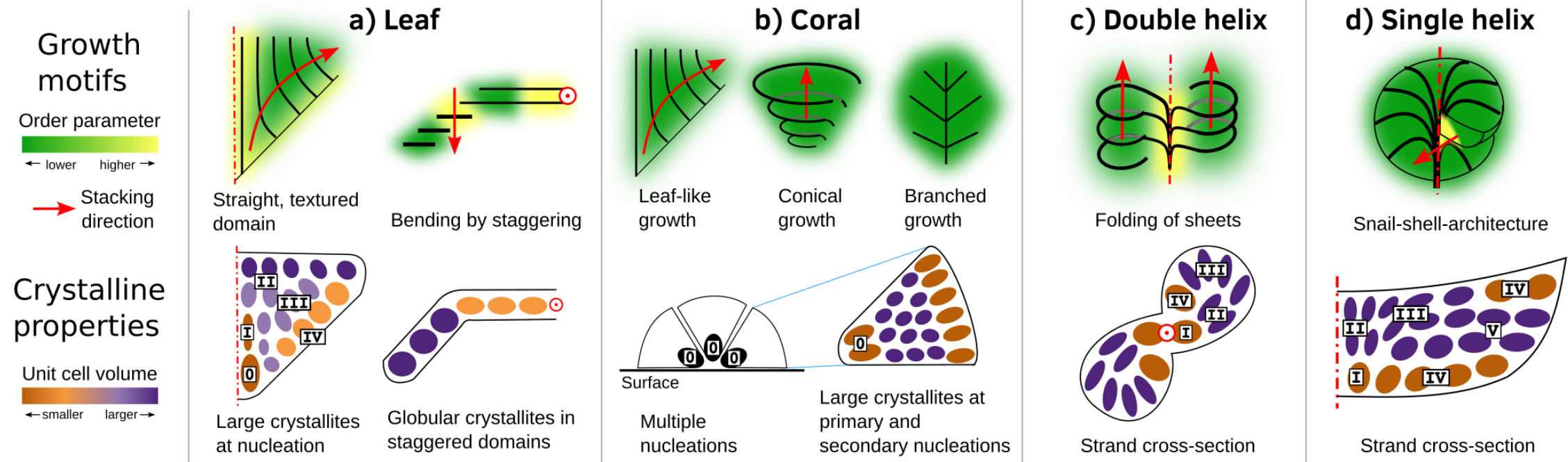

**Figure 8.** Summary of recurring structural motifs and crystalline property regimes: Top row: schematic summary of the dominant growth-related structural motifs inferred from the 3D orientation reconstructions for **(a)** leaf, **(b)** coral, **(c)** double helix, and **(d)** single helix morphologies. Black lines correspond to nematic directors derived from texture tomography, red arrows indicate the stacking direction of the crystallites. Bottom row: schematic crystalline-property regimes derived from texture tomography and diffraction tomography. The helical morphologies further exhibit recurring internal regions and related strand cross-sectional organization.

**Table 1.** Experimental details (Number of projections, voxel size, field of view and estimated Dose) for the biomorph samples

| Sample | No. of projections | Voxel size [µm] | Field of view [µm] | Dose [Gy] |
|---|---|---|---|---|
| Leaf | 343 | 0.50 | 60 x 53 | $3.70 \cdot 10^{10}$ |
| Coral | 136 | 0.25 | 60 x 62 | $1.47 \cdot 10^{10}$ |
| Double Helix with nucleus | 162 | 0.25 | 62.5 x 33.5 | $1.75 \cdot 10^{10}$ |
| Single helix | 194 | 0.50 | 90 x 100 | $2.10 \cdot 10^{10}$ |

# Supplementary material

**The crystalline properties of silica biomorphs vary within and between morphologies**

*Moritz P.K. Frewein[1*], Britta Maier[2*], Moritz L. Stammer[1], Isabella Silva Barreto[1], Anastasiia Sadetskaia[3], Asma Medjahed[4], Remi Tucoulou[4], Julie Villanova[4], Manfred Burghammer[4], Henrik Birkedal[3], and Tilman A. Grünewald [1‡]*

[1] Aix Marseille Univ, CNRS, Centrale Med, Institut Fresnel, Marseille, France,
[2] University of Konstanz, Konstanz, Germany

[3] Department of Chemistry and INANO, Aarhus University, Aarhus, Denmark

[4] European Synchrotron Radiation Facility, Grenoble, France

‡ E-mail tilman.grunewald@fresnel.fr
* these authors contributed equally

List of items:

Overview of reported formation conditions and morphology regimes of silica–witherite biomorphs

Discussion of attainable experimental resolution

Movie S1 Leaf

Movie S2 Coral

Movie S3 Double helix with nucleus

Movie S4 Single helix

Supplementary Figures S1 – S9

**Overview of reported formation conditions and morphology regimes of silica–witherite biomorphs**

According to literature, geological witherite (i.e., equilibrium-grown crystals expressing rhombic crystal facets), globular structures, and silica-witherite biomorphs form under distinct but somewhat overlapping conditions governed by temperature, pH, silicate and barium concentrations, $CO_2$ uptake, and the presence of additives (1–7). Because experimental protocols vary substantially, comparisons are largely qualitative. Here, we summarize published information concerning silica-witherite biomorph formation in solution, drawing mainly from the morphograms of Rouillard (1) and Opel (6), as well as multiple mechanistic studies (2–5, 8). Figure S1 summarizes these observations in two conceptual morphograms as a function of temperature and silicate concentration (Fig S1a) and pH and $BaCO_3$ supersaturation (Fig S1b), complemented by studies on the influence of additives (9–17) and reaction-diffusion models (18–20).

Reported synthesis conditions for silica-witherite biomorphs typically span temperatures of 20-50 °C, pH values of ~9-12, a silicate or dissolved silica concentration of ~7-18 mM, and a barium ion concentration of ~0.6-250 mM (1, 6, 9). At low or absent silicate concentrations, and at pH values outside the commonly reported biomorph window (~10.2-11.1), geological witherite crystals (I) are observed (8, 21, 22). Accordingly, reduced silicate-witherite coupling is expected, which disfavors biomorph-type nanocrystalline textures. As conditions approach the biomorph regime, globules and their aggregates (II) appear at suitable pH but too low temperature (6) or mismatched $Ba^{2+}$-silicate ratios (22) due to imbalanced precipitation and silicate polymerization.

Within the typical silica–witherite biomorph window—characterized by pH values of ~10.2–11.1, moderate temperature (20-25 °C), and dissolved silica concentrations of ~7–18 mM—curvilinear morphologies dominate, most prominently leaf-like sheets (III, Fig S1c) and helicoid-like structures (IV, Fig S1e, f) (4, 6, 8). Coral-like structures (V, Fig S1d) form at elevated temperatures (50 °C) (6), high supersaturation (6, 8), increased pH (3), or in the presence of multivalent ions or polymers (5). Funnel-like morphologies (trumpet-, vase-, or urn-shaped) (VI) are observed near the upper pH limit of the biomorph regime (~10.9-11.3)(1, 8) or as coral-like structures at elevated temperatures (50 °C) (6, 23).

Additives, including foreign ions, biomolecules, and polymers, modulate these morphologies by influencing reaction-diffusion fields, growth kinetics, and front stability, as demonstrated experimentally (9–16) and explained by reaction-diffusion models (18–20). Thus, they promote planar growth, branching, or stabilize curved fronts (4, 5, 9). The influence of additives on the silica–witherite biomorph formation is either by directly modifying barium carbonate nucleation and growth (9–11), by altering silicate oligomerization and condensation (5, 12, 14), or—most effectively—by perturbing the coupled reaction–diffusion processes that link both phases (13, 15, 20).

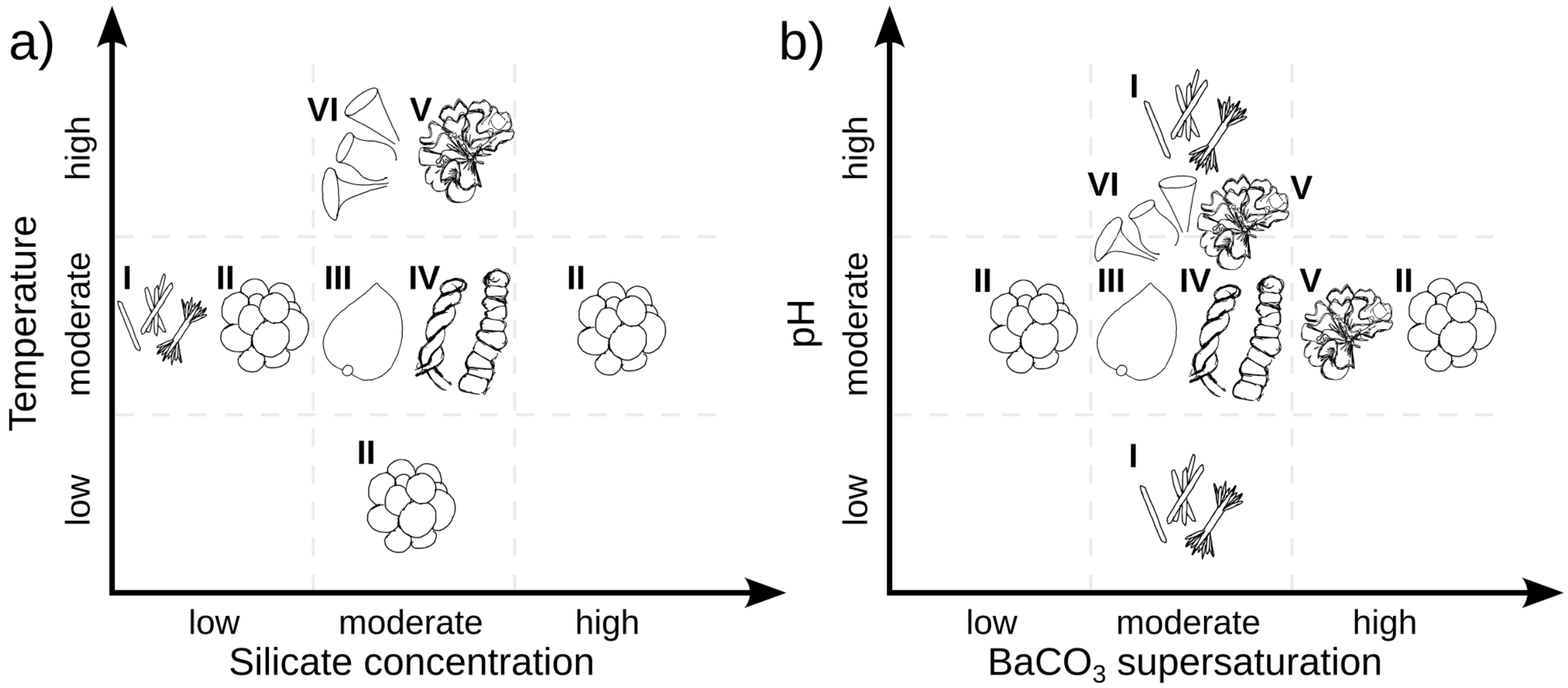

**Figure S1.** Conceptual morphogenetic maps illustrating how silica–witherite biomorph structures depend on temperature, silicate concentration (a), pH, and $BaCO_3$ supersaturation (b). Geological witherite (I) dominates outside the biomorph pH range or at low silicate concentration. A) Temperature–silicate concentration map. Approaching the biomorph regime yields globules (II) where carbonate precipitation and silicate polymerization are still unbalanced. In the 'biomorph window', leaf-like sheets (III) and helicoids (IV) arise at moderate temperature and silicate levels; coral-like shapes (V) and funnel-like structures (VI) form at higher temperature. B) pH– $BaCO_3$ supersaturation map. Within, morphology follows supersaturation: globules (II) at moderate to high, leaf-like structures (III) at intermediate, and helicoids (IV) at lower supersaturation. Coral-like shapes (V) arise at higher pH or high supersaturation, while funnels (VI) occur at an even higher pH near the upper pH boundary. Additives and multivalent ions shift these regimes by altering branching and curvature.

Although supersaturation is often considered a primary control parameter, the availability, transport, and speciation of silicate are equally important for silica-witherite biomorphs (5). $BaCO_3$ supersaturation varies strongly with pH but weakly with temperature (8), whereas silicate polymerization is highly sensitive to temperature, ionic strength, and time, increasing by more than an order of magnitude between 25 °C and ~60 °C (24, 25). Silica-witherite biomorph formation thus reflects the dynamic interplay between barium carbonate supersaturation and evolving silicate oligomers (5, 20).

Mechanistic and reaction-diffusion models indicate that distinct biomorph morphologies emerge from differences in the stability, curvature, and dynamics of the active growth front rather than from differences in crystallographic symmetry alone (10, 20). Within this framework, leaf-like morphologies are associated with relatively stable, weakly curved growth fronts sustained by moderate nucleation rates (1, 5), whereas helicoidal structures arise when the growth front undergoes sustained lateral deflection, consistent with oscillatory reaction–diffusion dynamics coupled to local silica accumulation (10, 19).
Trumpet- and coral-like morphologies can be interpreted as manifestations of directionally biased or intermittently blocked growth fronts (2, 18), leading to open, funnel-like or branched architectures.

**Discussion on the attainable experimental resolution**

The methods employed in this study (Texture tomography, XRD-CT coupled with Rietveld refinement) are able to unveil nanoscale crystalline properties. It is important to keep some kind of 'resolution limit' in mind, as far as they are established yet. With TexTOM being a very recently developed method, a systematic assessment concerning resolution limits is still a work in progress but a resolution in the order of 5° is a conservative estimate to start with and the orientation changes observed in the samples are substantially larger than this. For XRD-CT/Rietveld, a study on a nanocrystalline hydroxyapatite phantom sample (26) determined an accuracy in the order of 0.02% for the determination of lattice parameters and 5% for $V_{particle}$ based on the standard deviation of the investigated samples. In the context of this work, this means that the differences observed within and between the $V_{UC}$ on the order of ~0.5 $Å^3$/0.15% are meaningful as are the $V_{crystallite}$ changes in the order of 3000 $nm^3$/20%.

**Legend for Movie S1**

3D animation and slicing through the 3D volume of the morphology 'Leaf'. The glyphs and flow-lines represent the nematic director. The color scale corresponds to the order parameter.

**Legend for Movie S2**

3D animation and slicing through the 3D volume of the morphology 'Coral'. The glyphs and flow-lines represent the nematic director. The color scale corresponds to the order parameter.

**Legend for Movie S3**

3D animation and slicing through the 3D volume of the morphology 'Double helix' with a zoom on the nucleation zone. The glyphs and flow-lines represent the nematic director. The color scale corresponds to the order parameter.

**Legend for Movie S4**

3D animation and slicing through the 3D volume of the morphology 'Single helix'. The glyphs and flow-lines represent the nematic director. The color scale corresponds to the order parameter.

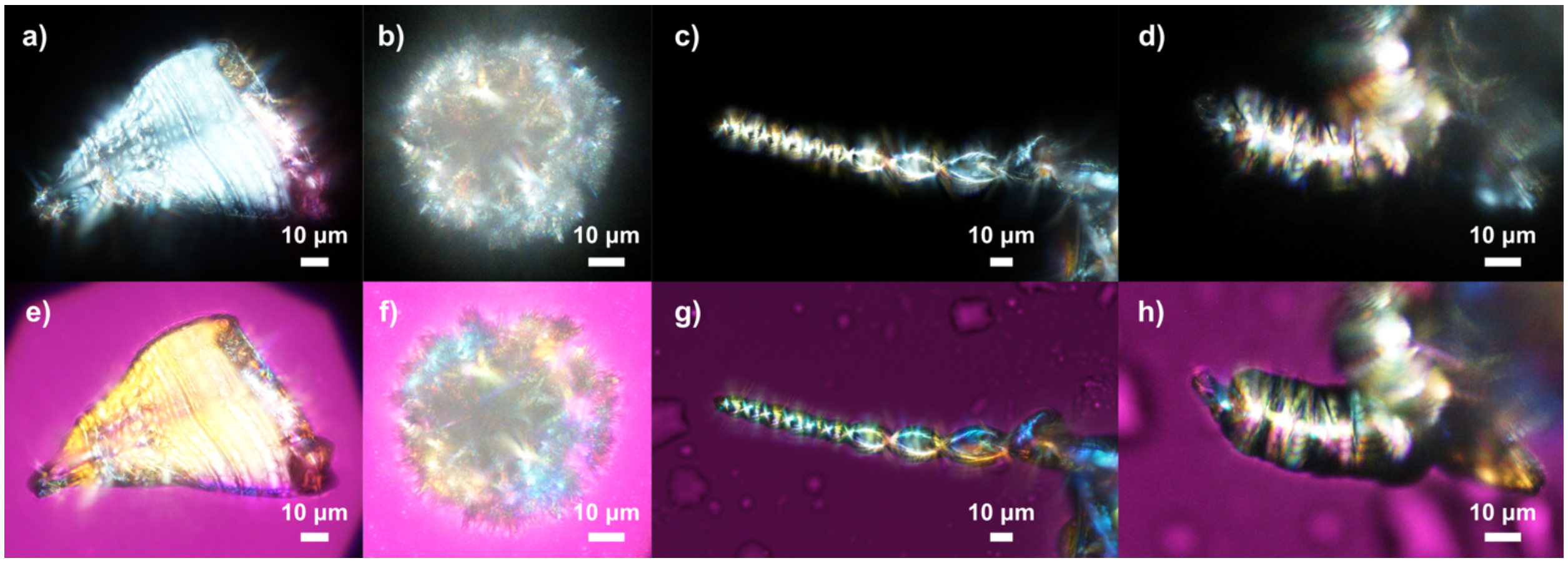

**Figure S2.** Optical microscopy images of representative samples from all biomorph morphologies under crossed polarizers. They were produced by a max-projection from 7 to 10 images at different focal heights across the samples. Subfigures a-d) show images from leaf, coral, double helix and single helix, respectively. Subfigures e-h) show the same images after insertion of a λ=530 nm retardance plate. Note the presence of 'spike'-like features in the coral structure, similar to the X-ray CT data.

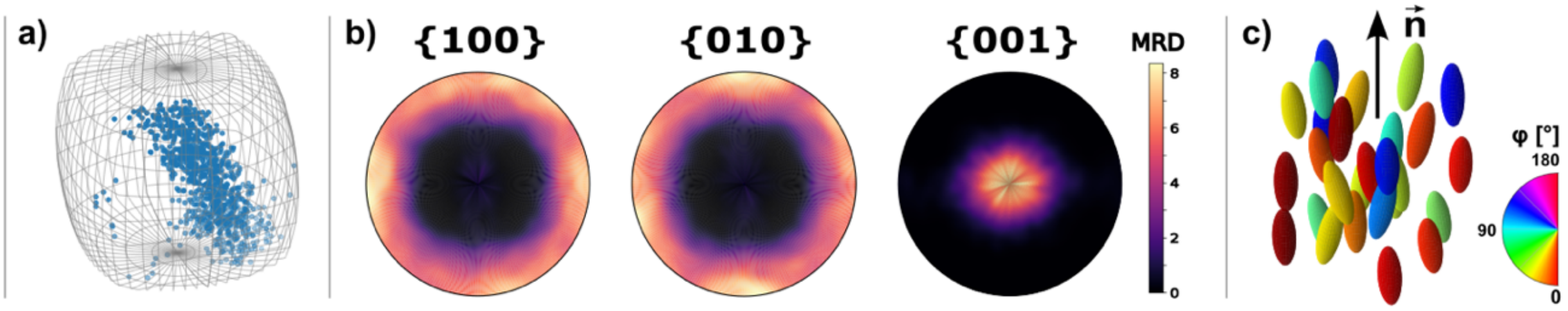

**Figure S3.** a) Representative texture of silica-witherite biomorphs represented as an ODF in axis-angle parametrization. Each point corresponds to one crystal orientation and the probability is given by the density of points. The orientation in each point is given by a rotation axis (defined by the unit direction from the origin to the respective point) and a rotation angle (distance from the origin). The wire-frame that marks the borders of the fundamental zone of orientation space (the sub-space that contains only unique crystal orientations for the 222-point group symmetry of witherite). The origin (no rotation) lies at the center of this zone. The column-shape of the distribution is the footprint of a fiber texture. For clarity, we added two other representations:
b) Visualization of the same texture by pole figures of the 3 orthogonal crystal axes in the orthorhombic crystal system. The probability of finding a given crystal direction is displayed in a stereographic projection in units of multiples of a random distribution (MRD). For better visibility, the most likely value of the distribution was rotated to the center of the fundamental zone before calculating the pole figures. The fiber texture can thereby easily be discerned, as one crystal axis – here the {001} or c-direction – has a narrow distribution around the center, while the other axes are approximately randomly distributed.
c) Real space visualization of a fiber texture by means of ellipsoidal crystallites, whose long axis corresponds to the {001} direction. The color-coded angle $\phi$ is the rotation angle around of the [1,0,0] axis around the long axis of the particle, which follows a flat distribution. The nematic director $\vec{n}$ marks the mean direction of the long axis of the particles.

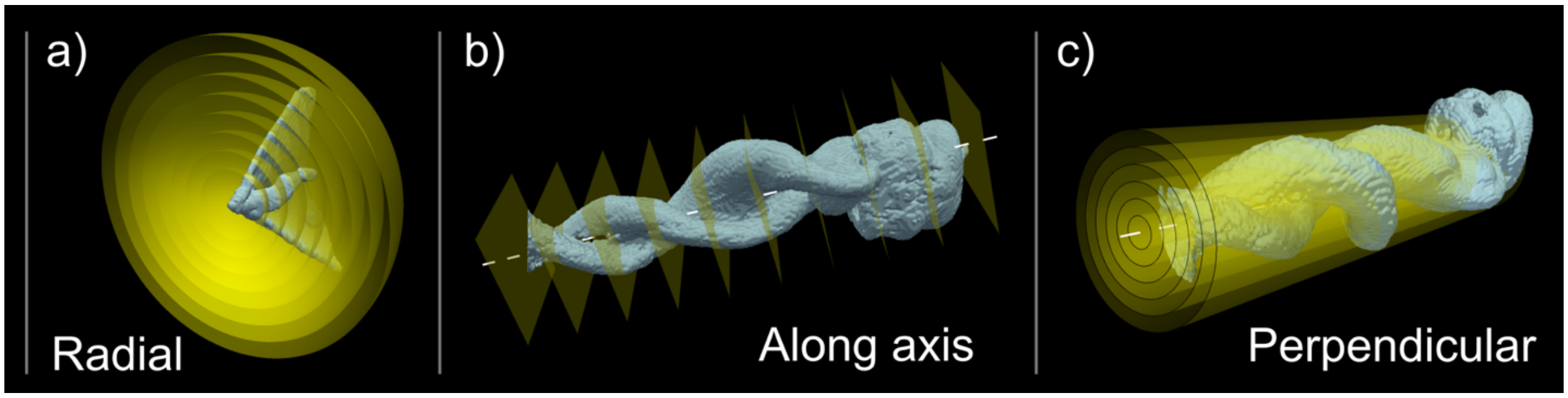

**Figure S4.** Integration schemes for trendlines through the biomorphs. The radial scheme in subfigure a) was used for leaf and coral morphologies. The data was divided into concentric shells around the nucleation center, and trendline means and standard deviations were calculated from the points lying within 2 shells, respectively. b) and c) show the schemes used for the helices: After determining the central axis of the helix, means and standard deviations were calculated between parallel planes normal to the axis (b), and between concentric cylinders around the axis (c).

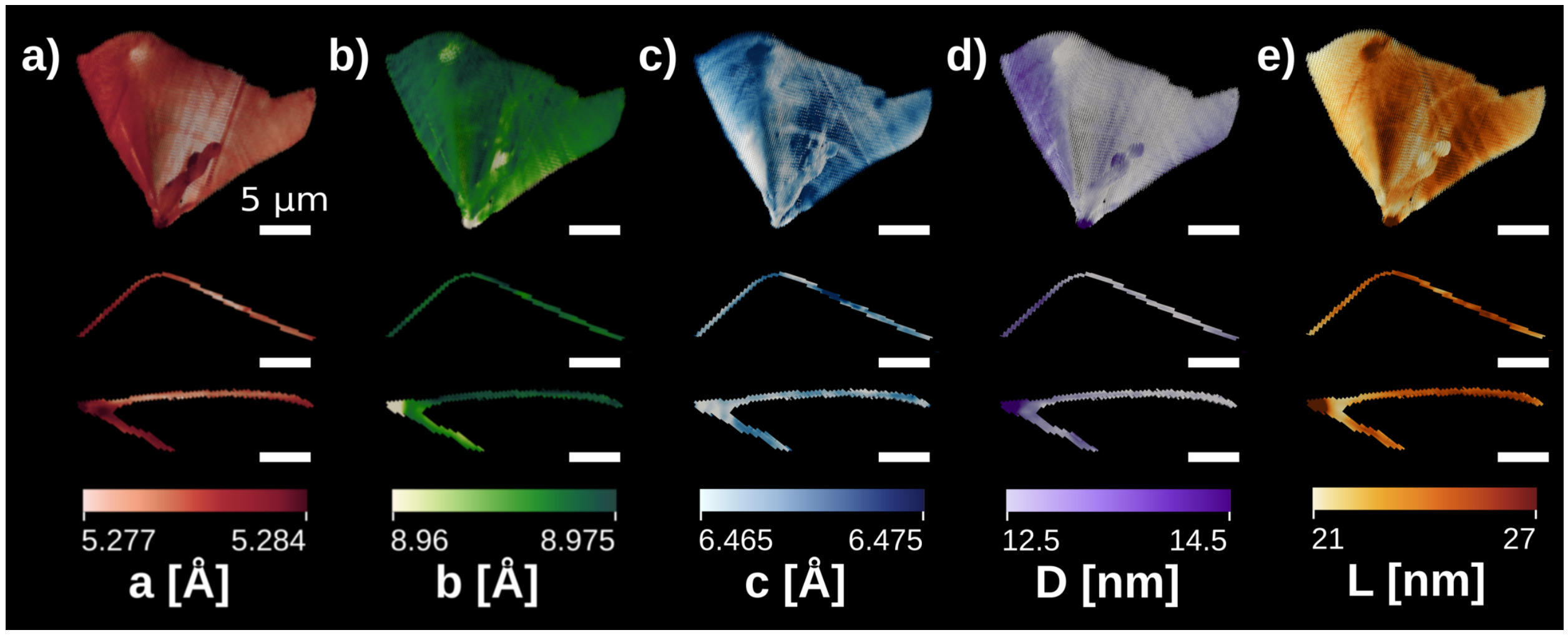

**Figure S5.** Crystalline properties of the leaf-like morphology: a-c) show the dimensions of the witherite unit cell. d) and e) are the diameter and length of the coherently scattering domains of the witherite nano-rods. All color scales are linear. A visualization of the morphology with the exact location of the cuts is shown in the main text, Fig. 3.

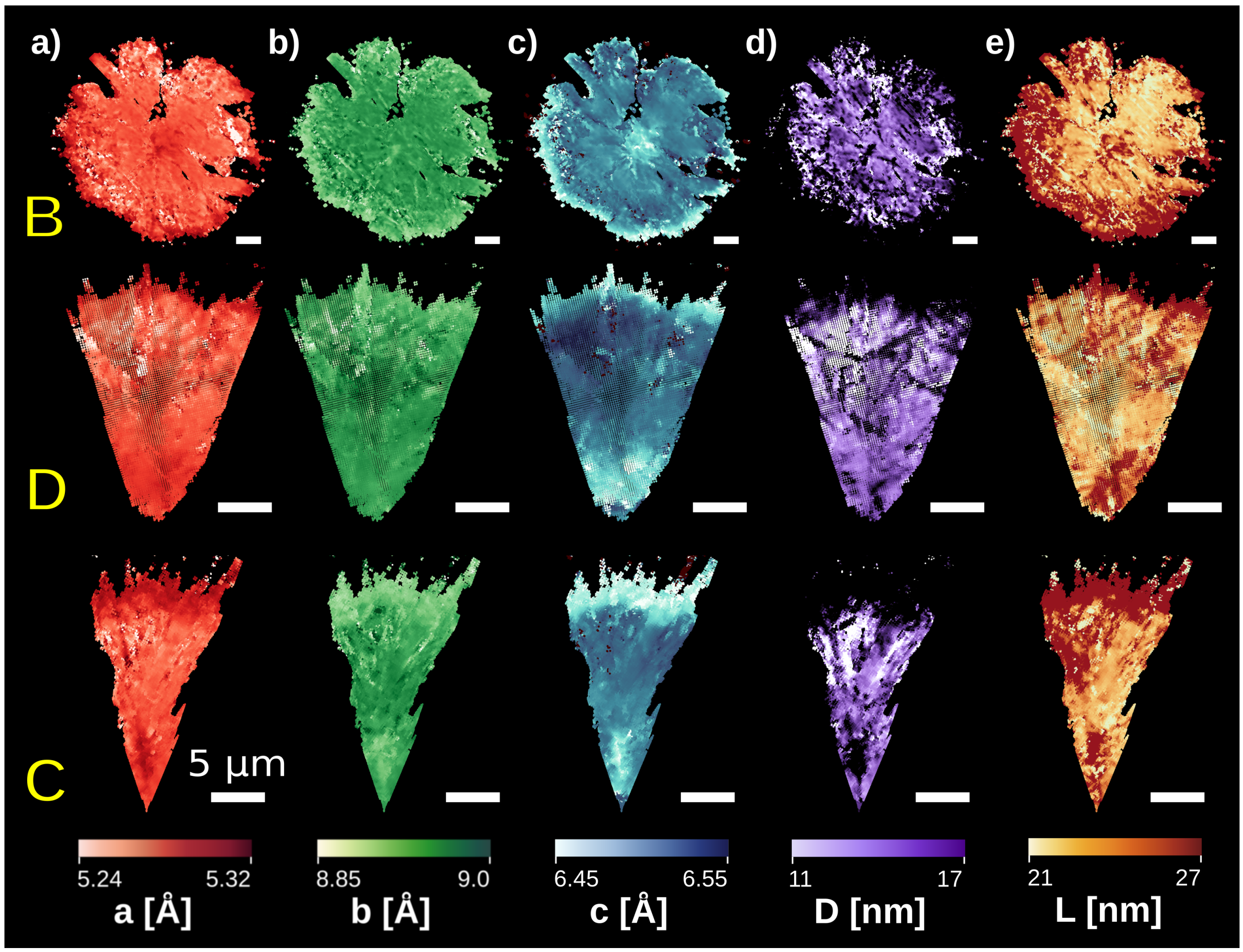

**Figure S6.** Crystalline properties of the coral morphology: a-c) show the dimensions of the witherite unit cell. d) and e) are the diameter and length of the coherently scattering domains of the witherite nano-rods. All color scales are linear. A visualization of the morphology with the exact location of the base-cut (B) as well as the L- and C-domains is shown in the main text, Fig. 4.

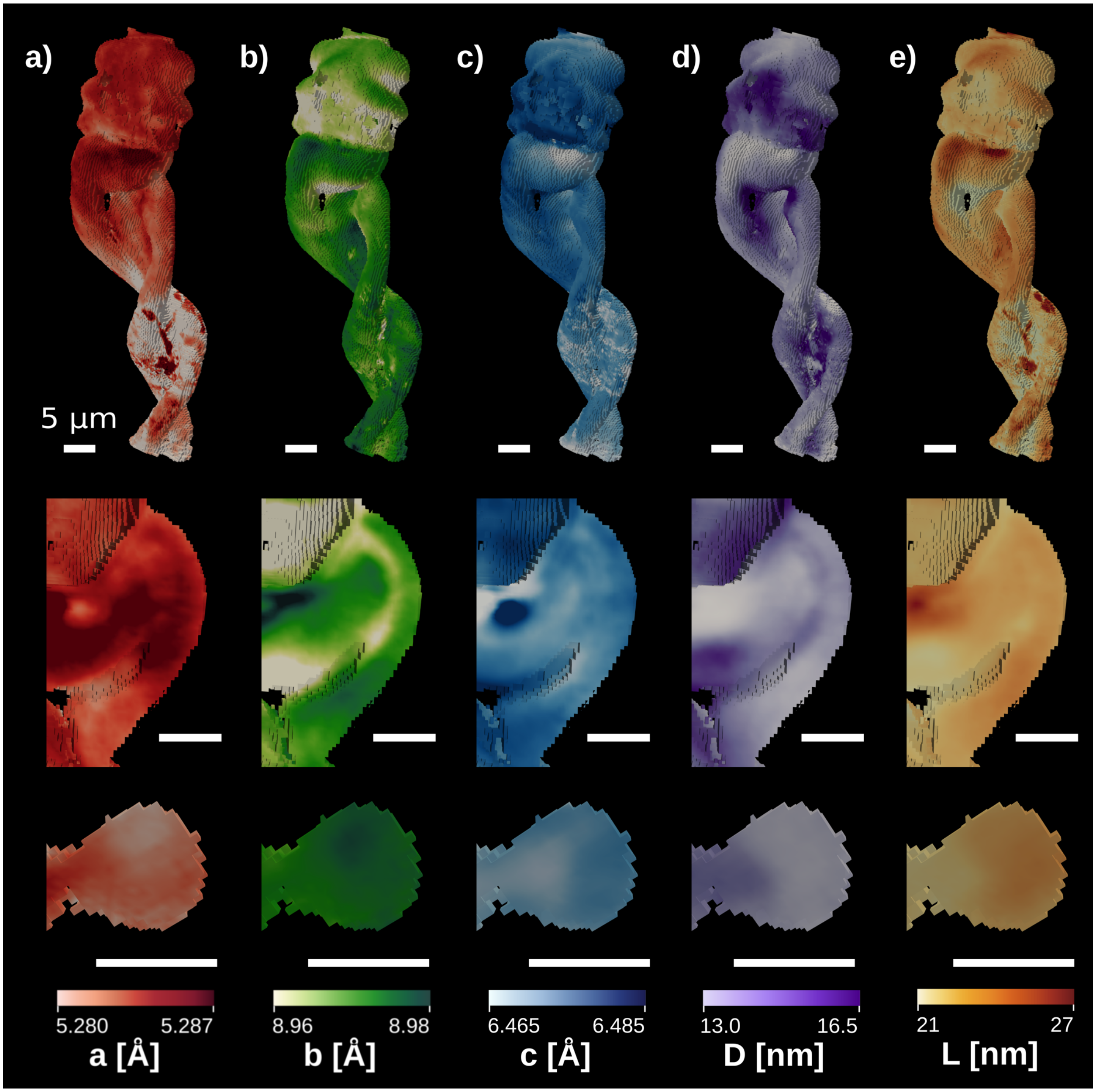

**Figure S7.** Crystalline properties of the double helix morphology: a-c) show the dimensions of the witherite unit cell. d) and e) are the diameter and length of the coherently scattering domains of the witherite nano-rods. All color scales are linear. A visualization of the morphology with the exact location of the cuts is shown in the main text, Fig. 7.

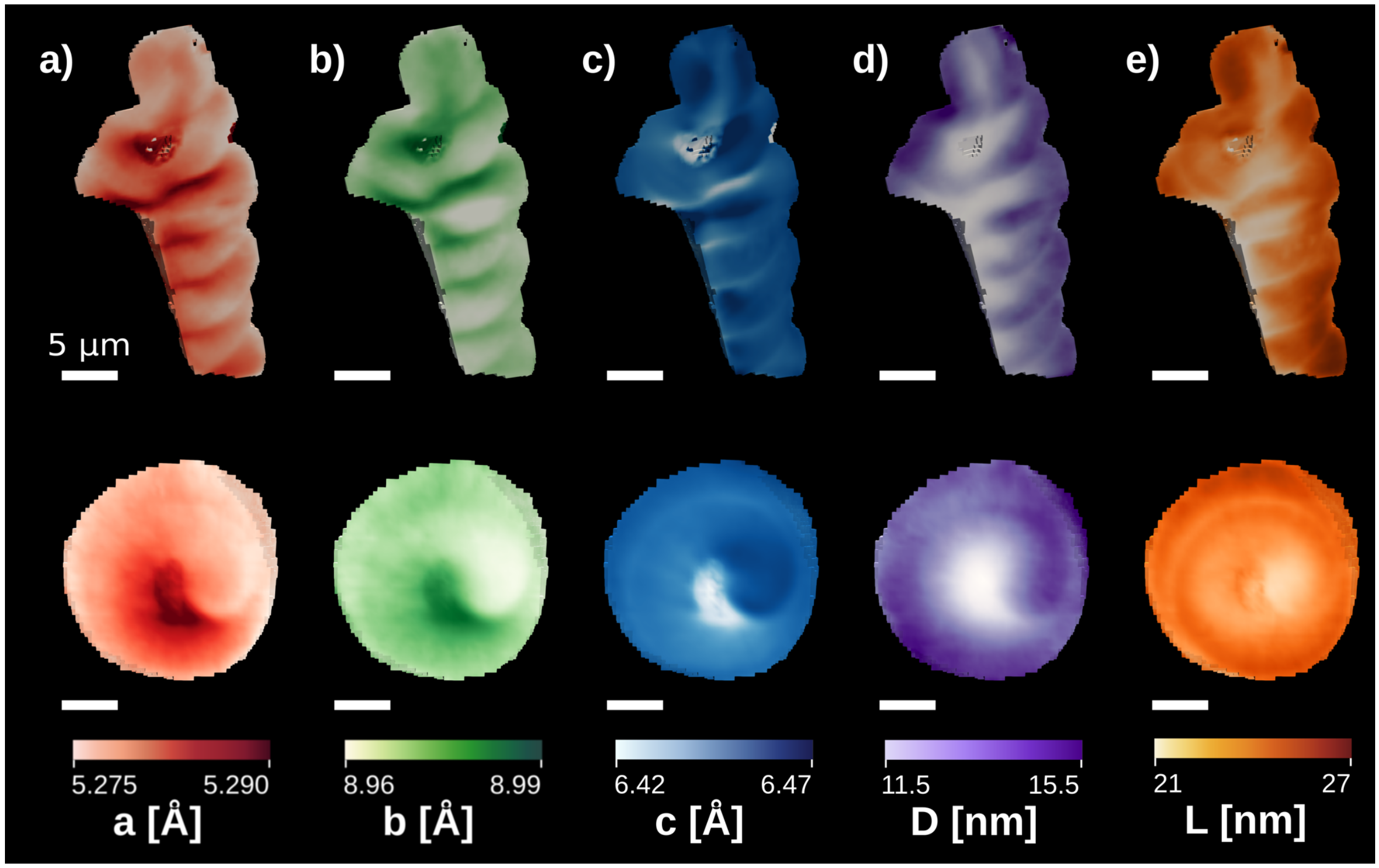

**Figure S8.** Crystalline properties of the single helix morphology: a-c) show the dimensions of the witherite unit cell. d) and e) are the diameter and length of the coherently scattering domains of the witherite nano-rods. All color scales are linear. A visualization of the morphology with the exact location of the cuts is shown in the main text, Fig. 7.

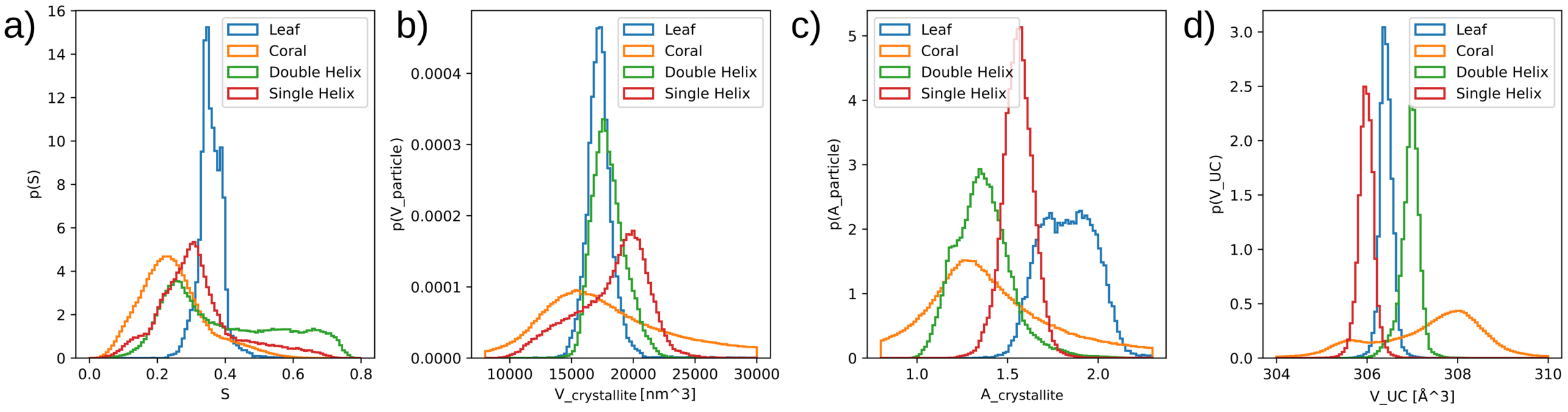

**Figure S9.** Histograms of crystalline properties for all four morphologies normalized to probability density functions. Subfigures a-d) respectively show the distributions of order parameter S, volume of coherently scattering crystalline domains $V_{crystallite}$, shape anisotropy of these domains $A_{crystallite}$, and volume of the witherite unit cell $V_{UC}$.

**References**

1. J. Rouillard, J. -M. García-Ruiz, J. Gong, M. A. Van Zuilen, A morphogram for silica-witherite biomorphs and its application to microfossil identification in the early earth rock record. *Geobiology* **16**, 279–296 (2018).

2. C. N. Kaplan, *et al.*, Controlled growth and form of precipitating microsculptures. *Science* **355**, 1395–1399 (2017).

3. W. L. Noorduin, A. Grinthal, L. Mahadevan, J. Aizenberg, Rationally Designed Complex, Hierarchical Microarchitectures. *Science* **340**, 832–837 (2013).

4. J. M. García-Ruiz, E. Melero-García, S. T. Hyde, Morphogenesis of Self-Assembled Nanocrystalline Materials of Barium Carbonate and Silica. *Science* **323**, 362–365 (2009).

5. M. Kellermeier, *et al.*, Growth Behavior and Kinetics of Self-Assembled Silica–Carbonate Biomorphs. *Chemistry A European J* **18**, 2272–2282 (2012).

6. J. Opel, *et al.*, Structural Transition of Inorganic Silica–Carbonate Composites Towards Curved Lifelike Morphologies. *Minerals* **8**, 75 (2018).

7. J. Eiblmeier, *et al.*, New insights into the early stages of silica-controlled barium carbonate crystallisation. *Nanoscale* **6**, 14939–14949 (2014).

8. J. Eiblmeier, M. Kellermeier, D. Rengstl, J. Manuel García-Ruiz, W. Kunz, Effect of bulk pH and supersaturation on the growth behavior of silica biomorphs in alkaline solutions. *CrystEngComm* **15**, 43–53 (2013).

9. M. Kellermeier, *et al.*, Additive-induced morphological tuning of self-assembled silica–barium carbonate crystal aggregates. *Journal of Crystal Growth* **311**, 2530–2541 (2009).

10. E. Nakouzi, R. Rendina, G. Palui, O. Steinbock, Effect of inorganic additives on the growth of silica–carbonate biomorphs. *Journal of Crystal Growth* **452**, 166–171 (2016).

11. S. R. Islas, M. Cuéllar-Cruz, Silica-Carbonate of Ba(II) and $Fe^{2+}$ /$Fe^{3+}$ Complex as Study Models to Understand Prebiotic Chemistry. *ACS Omega* **6**, 35629–35640 (2021).

12. N. Sánchez-Puig, *et al.*, The Influence of Silicateins on the Shape and Crystalline Habit of Silica Carbonate Biomorphs of Alkaline Earth Metals (Ca, Ba, Sr). *Crystals* **11**, 438 (2021).

13. E. Nakouzi, H. M. Fares, J. B. Schlenoff, O. Steinbock, Polyelectrolyte complex films influence the formation of polycrystalline micro-structures. *Soft Matter* **14**, 3164–3170 (2018).

14. M. Cuéllar-Cruz, S. R. Islas, G. González, A. Moreno, Influence of Nucleic Acids on the Synthesis of Crystalline Ca(II), Ba(II), and Sr(II) Silica–Carbonate Biomorphs: Implications for the Chemical Origin of Life on Primitive Earth. *Crystal Growth & Design* **19**, 4667–4682 (2019).

15. M. Cuéllar-Cruz, *et al.*, Formation of Crystalline Silica–Carbonate Biomorphs of Alkaline Earth Metals (Ca, Ba, Sr) from Ambient to Low Temperatures: Chemical Implications during the Primitive Earth's Life. *Crystal Growth & Design* **20**, 1186–1195 (2020).

16. N. Sánchez-Puig, *et al.*, Controlling the morphology of silica–carbonate biomorphs using proteins involved in biomineralization. *J Mater Sci* **47**, 2943–2950 (2012).

17. M. Cuéllar-Cruz, A. Moreno, Synthesis of Crystalline Silica–Carbonate Biomorphs of Ba(II) under the Presence of RNA and Positively and Negatively Charged ITO Electrodes: Obtainment of Graphite via Bioreduction of $CO_2$ and Its Implications to the Chemical Origin of Life on Primitive Earth. *ACS Omega* **5**, 5460–5469 (2020).

18. A.-K. Malchow, A. Azhand, P. Knoll, H. Engel, O. Steinbock, From nonlinear reaction-diffusion processes to permanent microscale structures. *Chaos: An Interdisciplinary Journal of Nonlinear Science* **29**, 053129 (2019).

19. P. Knoll, N. Wu, O. Steinbock, Spiraling crystallization creates layered biomorphs. *Phys. Rev. Mater.* **4**, 063402 (2020).

20. P. Knoll, E. Nakouzi, O. Steinbock, Mesoscopic Reaction–Diffusion Fronts Control Biomorph Growth. *J. Phys. Chem. C* **121**, 26133–26138 (2017).

21. M. Kellermeier, H. Cölfen, J. M. García-Ruiz, Silica Biomorphs: Complex Biomimetic Hybrid Materials from "Sand and Chalk." *Eur J Inorg Chem* **2012**, 5123–5144 (2012).

22. E. Bittarello, F. Roberto Massaro, D. Aquilano, The epitaxial role of silica groups in promoting the formation of silica/carbonate biomorphs: A first hypothesis. *Journal of Crystal Growth* **312**, 402–412 (2010).

23. M. Kellermeier, "Co-mineralization of alkaline-earth carbonates and silica," Universität Regensburg. (2011).

24. H. P. Rothbaum, A. G. Rohde, Kinetics of silica polymerization and deposition from dilute solutions between 5 and 180°C in *Journal of Colloid and Interface Science*, (1979), pp. 533–559.

25. O. Weres, A. Yee, L. Tsao, Kinetics of silica polymerization. *Journal of Colloid and Interface Science* **84**, 379–402 (1981).

26. S. Frølich, *et al.*, Diffraction tomography and Rietveld refinement of a hydroxyapatite bone phantom. *J Appl Crystallogr* **49**, 103–109 (2016).